\documentclass{aa}  
\usepackage[varg]{txfonts}
\usepackage{graphicx}
\usepackage{xcolor}       
\usepackage{placeins}
\usepackage[colorlinks=true,
            linkcolor=blue,
            citecolor=blue,
            urlcolor=blue]{hyperref}

\newcommand{\MS}{\ifmmode{\,}\else\thinspace\fi{\rm M}\ifmmode_{\odot}\else$_{\odot}$\fi}
\newcommand{\LS}{\ifmmode{\,}\else\thinspace\fi{\rm L}\ifmmode_{\odot}\else$_{\odot}$\fi}
\newcommand{\RS}{\ifmmode{\,}\else\thinspace\fi{\rm R}\ifmmode_{\odot}\else$_{\odot}$\fi}
\newcommand{\mesa}{\texttt{MESA}}
\newcommand{\MESA}{\mesa}
\newcommand{\rsp}{\texttt{RSP}}
\newcommand{\RSP}{\rsp}
\newcommand{\gyre}{\texttt{gyre}}
\newcommand{\fcor}{\ensuremath{f_{\rm H}}}
\newcommand{\fenv}{\ensuremath{f_{\rm env}}}
\newcommand{\chis}{\ensuremath{\chi^2}}
\newcommand{\Teff}{\ensuremath{T_{\rm eff}}}  
  
\defcitealias{Pilecki-2018}{P18}

\begin{document}

\title{Pulsation periods reveal tension between theoretical and empirical radii for classical Cepheids in eclipsing binary systems}

\titlerunning{Tension between theoretical and empirical radii for classical Cepheids}

\author{O. Zi\'{o}\l{}kowska\inst{1}\email{oliwiakz@camk.edu.pl}
\and R. Smolec \inst{1}\corrauth{smolec@camk.edu.pl}
\and R. Singh Rathour\inst{2}\email{rajeev.rathour@oca.eu}
\and V. Hocd\'{e}\inst{1}\email{Vincent.Hocde@oca.eu}
}

\institute{Nicolaus Copernicus Astronomical Center, Polish Academy of Sciences, Bartycka 18, 00-716 Warszawa, Poland
\and Université C\^{o}te d'Azur, Observatoire de la C\^{o}te d'Azur, CNRS, Laboratoire Lagrange, France}

\date{Received date /
Accepted date}

  \abstract
   {With their precisely determined physical parameters, classical Cepheids in eclipsing binary systems are often used to constrain stellar evolution and pulsation theories. To this end, their position in the Hertzsprung–Russell diagram, effective temperature and luminosity, and radius are commonly used when matching best-fitting evolutionary models. However, the pulsation period of classical Cepheids, the most precise observable, is rarely used in such studies.}
   {We explore how including the pulsation period as a constraint in matching evolutionary models affects the best-fitting solution. As the pulsation period follows the period–mean density relation, we examine whether it provides information consistent with that based on the stellar radius.} 
   {We modeled four eclipsing binary systems with classical Cepheids from the Large Magellanic Cloud. We used $\chis$ minimization to find the best-matching evolutionary model from a grid of models computed with Modules for Experiments in Stellar Astrophysics (\MESA). The evolutionary models are supplemented with pulsation periods computed with the Radial Stellar Pulsation (\RSP) module of \MESA, with nonlinear period corrections taken into account.}
   {Depending on whether the radius or the pulsation period is used to select the best-fitting model, discrepant solutions are obtained. In solutions selected based on the pulsation period, the stellar radius is systematically too low compared with observations. Conversely, for solutions based on the radius, the pulsation period is systematically too long. The tension amounts to a few sigma for stars with precisely determined radii, and part of it is traced to a nonlinear increase in radius for large-amplitude pulsators, which has not been studied in detail in the literature.}
   {When using classical Cepheids in eclipsing binary systems to constrain stellar models, we recommend using the pulsation period instead of the radius. A systematic study of nonlinear effects on the stellar radius in large-amplitude pulsators is needed.}


\maketitle
\nolinenumbers

\section{Introduction}\label{sec:intro}

Classical Cepheids (or type I Cepheids; hereafter Cepheids) are young intermediate-mass ($\sim$3--15\MS), radially pulsating variable stars with high amplitudes and typical pulsation periods of 1--100\,d. On the Hertzsprung-Russell diagram (HRD) they reside in a well-defined, narrow region, the classical instability strip (IS), where they are unstable to pulsations due to $\kappa$/$\gamma$ mechanism. Most Cepheids pulsate in either the fundamental (F) or first overtone (1O) mode. The second overtone and various combinations of double- and triple-mode pulsations have also been observed \citep[e.g.,][]{Soszynski-2015b}. From an evolutionary perspective, a stellar track can cross the IS three times. The first crossing happens during a rapid transition between the main sequence (MS) and the red giant branch (RGB), when energy is generated in a hydrogen burning shell. The second and third crossings happen during a more stable core-helium burning (CHeB), on the blue loop. Due to timescales, a majority of Cepheids are expected on the second and third crossings. For reviews, see \citet{Catelan-2015, Bono-2024}.

The well-defined relation between the absolute magnitudes and pulsation periods of Cepheids \citep[period-luminosity relation, the Leavitt law,][]{Leavitt-1907, Leavitt-1912} makes them powerful standard candles for determining distances in the local Universe \citep[e.g.,][]{Riess-2019, Freedman-2001}. They have been used as tracers of young stellar populations in galactic archaeology studies \citep[e.g.,][]{Skowron-2019, Jacyszyn-2016, Jacyszyn-2020, DeSomma-2025}. Another important use of these stars is testing the stellar evolution and pulsation theories \citep[e.g.,][]{Bono-1999, MoskalikDziembowski-2005, DeSomma-2024,Deka-2025}.

Upon comparing Cepheid mass estimates from stellar evolution and pulsation models, a discrepancy was revealed \citep[e.g.,][]{Christy-1968, Stobie-1969a, Stobie-1969b}. The difference, at first as high as 30--40\% \citep{Cox-1980}, was decreased by using revised OPAL opacities \citep[see][]{Iglesias-1991,Moskalik-1992}. Nowadays, it is predicted that the evolutionary masses of Cepheids are 17$\pm$5\% higher than the pulsation masses \citep{Keller-2008}. A particularly important role in studying the Cepheid mass discrepancy is played by Cepheids in detached, double-lined eclipsing binary (DLEB) systems, several of which have been identified in the Optical Gravitational Lensing Experiment (OGLE) data \citep{Soszynski-2008, Soszynski-2012, Udalski-2015b}.

\object{OGLE-LMC-CEP-0227} (hereafter, CEP-0227) was the first classical Cepheid observed in a well-detached DLEB, for which physical parameters were precisely measured, with a dynamical mass found at an unprecedented 1\% precision \citep{Pietrzynski-2010}. The measured mass of the Cepheid was in agreement with stellar pulsation theory, which proved that evolutionary models need to be refined.

The parameters of CEP-0227 were later improved in \citet{Pilecki-2013} and \cite{Pilecki-2018} (hereafter \citetalias{Pilecki-2018}). In particular, the mass of the companion turned out to be lower than the Cepheid mass (in contrast to original determination) after including more radial velocity measurements. CEP-0227 was modeled based on the initial parameters by \citet{Cassisi-2011} and \citet{Prada-Moroni-2012}. Using models with convective core overshooting and moderate mass loss, both studies were able to reproduce masses and radii of both components with equal ages. In another study, \citet{Neilson-2012} compared canonical models of CEP-0227 (without overshooting, rotation, and mass loss) with two alternative models. In the first, both stars were located on the Hertzsprung gap and had standard -- Large Magellanic Cloud (LMC)-like -- metallicity. In the second, the Cepheid was on a blue loop and had nonstandard metallicity. The authors found that alternative solutions cannot be ruled out and more information about the system is needed. Nonlinear pulsation modeling of CEP-0227 was conducted by \cite{Marconi-2013} who satisfactorily reproduced the observed light and radial velocity curves.

The next DLEB with a Cepheid that was used to determine the dynamic masses was \object{OGLE-LMC-CEP-1812} \citep[][hereafter CEP-1812]{Pietrzynski-2011}. The two stars were identified in a similar evolutionary stage and have significantly different masses. \citet{Neilson-2015} argued that the Cepheid is the product of a stellar merger of two MS stars. They also concluded that CEP-1812 might be an anomalous Cepheid or a link between classical and anomalous Cepheids and that it is not a reliable star for calibrating the Leavitt law or resolving the mass discrepancy problem. Recent discovery of another similar system, OGLE-LMC-CEP-1347 \citep{Pilecki-2022, Espinoza-2025}, might indicate that merger-origin Cepheids are more common.

A third system, \object{OGLE-LMC-CEP-4506} (hereafter CEP-4506; initial ID LMC562.05.9009) with dynamical mass derived by \citet{Gieren-2015}, had both components with similar masses, radii, and effective temperatures. The primary component pulsates in the fundamental mode, whereas the secondary component remains non-pulsating despite its position within the IS. It was suggested that future temperature estimates might place it just outside the red edge of the IS. However with physical parameters improved in \citetalias{Pilecki-2018}, the location of the non-pulsating component within the IS holds.

A DLEB system with two 1O Cepheids, \object{OGLE-LMC-CEP-1718} (hereafter CEP-1718), was analyzed by \citet{Gieren-2014}, but the results were not reliable due to observational issues, and an improved model was later published by \citetalias{Pilecki-2018}. This system only showed one shallow eclipse, so the determined physical parameters are not as precise as for F-mode Cepheids. In \citet{Pilecki-2015} another 1O Cepheid in a DLEB, \object{OGLE-LMC-CEP-2532} (hereafter CEP-2532) was analyzed, and dynamical masses were determined and later refined by \citetalias{Pilecki-2018}.

The presented sample of six Cepheids in five DLEBs has been used in several studies of the mass discrepancy problem. \citet{Anderson-2014, Anderson-2016} argued that including rotation and considering the crossing number suffices to fix the evolutionary masses of Cepheids. \citet{Miller-2020}, on the other hand, argues that rotation and overshooting alone are insufficient to model the distribution of period change rates of Cepheids and that another mechanism, such as pulsation-driven mass loss, is required \citep{Neilson-2008, Neilson-2009, Neilson-2009b, Neilson-2010}. Modeling studies of this sample of stars, after the updated solutions were published in \citetalias{Pilecki-2018}, are scarce. \citet{Deka-2025} compared evolutionary models with the sample of \citetalias{Pilecki-2018}, but their primary goal was to test an application of a turbulent convection model of \cite{Kuhfuss-1986}, as developed for GARSTEC evolutionary code by \cite{Ahlborn-2022}.

In our previous work \citep{Ziolkowska-2024, Ziolkowska-2026, Smolec-2026}, we explored evolutionary \MESA{} models for intermediate-mass stars, in particular Cepheids. We considered 2–8\MS{} models spanning a range of metallicities and overshooting parameters. In \cite{Ziolkowska-2024}, we established a reference model -- which we also adopted in the present study -- with a consistent set of assumptions on the solar metal mixture, convective boundaries, nuclear networks, and atmosphere models, and we examined its numerical convergence. We then varied these assumptions one at a time to assess their impact on evolutionary tracks (effective temperatures, \Teff{}; luminosities, $L$; and ages) at different evolutionary stages, and we estimated the associated uncertainties from the average differences relative to the reference model. In \citet{Ziolkowska-2026}, we analyzed the corresponding changes in surface abundances and the central C/O ratio, finding that abundances are a robust outcome of evolutionary models. In \citet{Smolec-2026}, we studied in detail the effects of metallicity, core and envelope overshooting, and mass loss on Cepheid evolutionary and pulsation properties.

In the present paper, we build on this work to study classical Cepheids in eclipsing binary systems. In previous studies, best-fitting evolutionary models were typically constrained using various combinations of parameters, including \Teff{}, $L$; radius, $R$; or surface gravity, $\log g$ \citep{Cassisi-2011, Prada-Moroni-2012, Neilson-2012, Neilson-2015, Deka-2025}. Our primary goal was to exploit the most precise observable -- the pulsation period -- and to examine how it affects the solutions. Since, for radial oscillations, the pulsation period is tightly linked to the stellar radius through the period–mean density relation, we tested whether solutions based on stellar radius and pulsation period are mutually consistent.

The structure of the paper is as follows. In Sect.~\ref{sec:methods} we summarize the known information on the sample of Cepheids in DLEBs, and we introduce our modeling tools and assumptions. In Sect.~\ref{sec:results} we present the results of our modeling. In Sect.~\ref{sec:discussion} we discuss our results.

\section{Methods}\label{sec:methods}

\subsection{Data and observational constraints for DLEBs}\label{ssec:data}

For convenience, we have collected the most recent data on Cepheids in DLEBs from \citetalias{Pilecki-2018} in Tab.~\ref{tab:dlebs} in the Appendix. A basic assumption for stellar binaries is common evolution of both components originating from the same material, and thus they are assumed to share the same metallicity and age. Mass is one of the primary diagnostics of evolutionary stage: the more massive star evolves faster and, in general, should be more luminous. This is not the case for the binary Cepheid system CEP-1718, which is therefore challenging to model under the assumption of identical composition and overshooting. 

A mass ratio close to unity implies that both stars should evolve at nearly the same rate, which holds for all systems except CEP-1812, where the primary appears significantly younger than the secondary. Modeling by \cite{Neilson-2015} suggests that the Cepheid is a product of a stellar merger; consequently, the system cannot be treated as a pair of coeval, well-detached stars. For this reason, we do not attempt to model the system directly. The merger scenario implies a nonstandard evolutionary history, violating the assumption of coeval evolution that underlies our binary modeling approach. Nevertheless, once formed, the merger product evolves according to the standard equations of stellar evolution and is currently observed as an ordinary Cepheid. We therefore make use of its accurately determined physical parameters later in this work to examine the role of the pulsation period in constraining evolutionary models. In particular, CEP-1812 serves as an independent consistency check supporting the conclusions drawn from the analysis of the two other F-mode Cepheids in eclipsing binary systems regarding the radius discrepancy.

Without spectroscopic metallicity determinations for our sample, there are no strict constraints on metallicity. The previous studies of Cepheids in DLEBs relied on various mean metallicity determinations for the LMC. \citet{Hocde-2023} list literature sources of metallicity determinations for Magellanic Clouds, based on supergiants \citep{Russell-1989,Hill-1995, Andrievsky-2001, Urbaneja-2017,Romaniello-2022}, which after averaging roughly give [Fe/H]=$-0.35$, equivalent to $Z$=0.006. We take this value as a reference metallicity in our modeling. Based on their photometric metallicity determination, \cite{Hocde-2023} finds that metallicity distribution of Cepheids in the LMC is narrow ($\sigma$=0.11dex). Nevertheless, this reference metallicity should be regarded only as a working assumption and not as a system-specific constraint. To account for the uncertainty in metallicity, we subsequently explore a range of metallicity values for each system.

We note, that modeling is affected by a degeneracy between metallicity and convective overshooting (mixing processes in general), as decreasing the metallicity and increasing overshooting, leads to brighter evolutionary tracks \citep[][]{Constantino-2018, Smolec-2026, Smolec-2026b}. Consequently, the lack of spectroscopic metallicity determinations for Cepheids in eclipsing binary systems constitutes a major limitation on fully exploiting the potential of these systems to constrain and test stellar evolution models.

Evolutionary period change rate (PCR) is an indicator of the evolutionary stage thanks to the period-mean density relation, $P\sim\rho^{-1/2}$ \citep{Ritter-1879}. When a Cepheid is evolving red-ward, on the first, and third crossings, the star expands, its mean density decreases, and period increases. On the second crossing the situation is reversed and the PCR is negative. 

Using OGLE data, \citetalias{Pilecki-2018} measured CEP-1718B's PCR to be  $\dot{P}=-18\pm 3\times10^{-9}$, implying that the star is likely on the second crossing. The other Cepheid in this system (CEP-1718A) and CEP-1812 appear to have constant periods according to \citetalias{Pilecki-2018}. We note that the determinations of PCR for these stars are based on OGLE data, which span a relatively short time baseline ($\sim$20\,yr). On such timescales, non-evolutionary period changes may mimic evolutionary ones, as investigated by \cite{Rathour-2025}. Consequently, we treat the negative PCR for CEP-1718B only as an indication that the star is on the second crossing, rather than as a strict constraint.

CEP-2532 was classified by \citet{Rathour-2025} as having a constant period within considered time span. For the remaining Cepheids in the sample, we are not aware of published PCR determinations.

\subsection{Evolutionary and pulsation tools}\label{ssec:evpuls}

For evolutionary and pulsation models we use Modules for Experiments in Stellar Astrophysics, \MESA{} \citep{Paxton-2011, Paxton-2013, Paxton-2015, Paxton-2018, Paxton-2019, Jermyn-2023}, version r-21.12.1. We provided a detailed description and studied numerical convergence of \MESA{} models for intermediate-mass stars in \citet{Ziolkowska-2024} and summarize the physical setup below. In \cite{Smolec-2026} we have demonstrated that evolutionary tracks computed with \MESA{} r-21.12.1 are fully consistent with those computed with more recent \MESA{} version.

We evolve our models from zero-age main sequence (ZAMS) till the end of the CHeB. For a given initial metallicity, $Z$, helium content, $Y$, is calculated assuming primordial helium abundance $Y_{\rm p}$=0.2485 from \cite{Komatsu-2011}, and helium enrichment $\Delta Y/\Delta Z$=1.5 \citep[see Eqs.~1--3 in][]{Ziolkowska-2024}. Distribution of heavy elements is scaled with the solar composition from \citet{Asplund-2009} (A09). Opacity is interpolated from OPAL tables \citep{Iglesias-1993,Iglesias-1996} at low temperature supplemented with \cite{Ferguson-2005} tables. Type 2 tables, that include enhanced C and O abundances are used during the CHeB. Atmosphere boundary condition comes from PHOENIX tables \citep{Hauschildt-1999a, Hauschildt-1999b} and \citet{Castelli-2003} models. Atmosphere and opacity tables are built into \MESA. We use mixing length theory (MLT) as formulated by \citet{Henyey1965} with $\alpha_{\rm MLT}$=1.77, which was obtained by calibration of solar model in Sect.~2.3 of \citet{Ziolkowska-2024}. We use Schwarzschild criterion for locating the convective boundaries together with predictive mixing scheme introduced in \MESA{} in \citet{Paxton-2018}. The equation of state comes from various tables and their blends: OPAL \citep{Rogers-2002}, SCVH \citep{Saumon-1995}, HELM \citep{Timmes-2000}, PC \citep{Pothekin-2010} and Skye \citep{Jermyn-2021}, depending on temperature, density, and composition. We use exponential prescription for overshooting \citep{Herwig-2000} from the convective core during the MS phase, and from the convective envelope on RGB. Corresponding overshooting efficiency parameters are denoted as \fcor{} and \fenv. We neglect overshooting from the helium-burning core. We assume mass loss on the RGB with formulation from \citet{Reimers-1975} (see Sect.~\ref{secapp:massloss}), and on the AGB with a formula from \citet{Blocker-1995}. In this work, we neglect rotation. The \MESA{} parameter file (\texttt{inlist}) that we used to compute evolutionary tracks is the same as in the Appendix in \cite{Ziolkowska-2024} (with stellar and overshooting parameters adjusted accordingly to modeled systems).

To calculate pulsation periods, we use the Radial Stellar Pulsations module \citep[\RSP;][]{Paxton-2019, Smolec-2008} incorporated in \MESA. In this work, we construct envelope models based on parameters coming from the evolutionary tracks (mass $M$, $L$, $\Teff$, hydrogen and metal content $X$, $Z$). The same microphysical assumptions, in particular EOS and opacities, are used for the pulsation as for the evolutionary models. 

The envelope consists of 200 (\texttt{nz}) Lagrangian mass cells and extends to temperature \texttt{T\_in} = $2\times 10^6$K. The outer 60 (\texttt{nzouter}) cells have constant mass, down to the anchor zone placed at \texttt{T\_anch} = $11\times 10^3$K, below which the cell mass increases geometrically inward \citep[see fig.~2 in][]{Paxton-2019}. \RSP{} adopts time-dependent convection–pulsation coupling, following the one-equation model of \citet{Kuhfuss-1986} with free parameters described in \citet{Paxton-2019} (their Tab.~4). In particular, set A is used throughout this work. This setup defines our default pulsation configuration. In Section~\ref{ssec:puls} we investigate how sensitive the pulsation periods are to the adopted grid as well as to other parameters, including different convective parameter sets and the adopted reference solar composition. The IS plotted in the figures of this study corresponds to hot IS as defined in \cite{Smolec-2026} (their tab. 5).

\subsection{Two model grids}\label{ssec:grids}

We employ two grids of stellar evolution models: an initial grid and a final grid. These grids differ in their underlying physical assumptions, the range of parameters explored, and the resolution of the parameter space (i.e., the spacing between grid points).

The initial grid is used to investigate the sensitivity of the results to the adopted grid parameters and to identify preliminary solutions. Based on these findings, a tailored final grid is constructed for each system considered. This refined grid is then used to derive the final solutions and to form the basis for the discussion of the results.

The initial model grid is constructed as follows. For each star, one value of mass is assumed, corresponding to the observed mass from Tab.~\ref{tab:dlebs}. Mass loss is neglected. Models are computed for nine metallicity values centered on $Z$=0.006, with $dZ$=0.001 step, from $Z$=0.002 to $Z$=0.010. We use four values for efficiency of overshooting from the convective core, \fcor=(0.00,\, 0.01,\, 0.02,\, 0.03), and three values for efficiency of overshooting from the convective envelope, \fenv=(0.00,\, 0.02,\, 0.04).  

Based on the initial solutions identified with the initial model grid, we construct final grids tailored to each of the systems. Their construction is described later in Sect.~\ref{ssec:gridone}. Here, we only note the main differences. The final model grid includes mass loss. As we argue later in Sect.~\ref{ssec:gridone}, overshooting from the convective envelope can be fixed and thus removed from the explored parameter space. We explore variations in mass within the observational uncertainties and increase the grid resolution.

Considering mass loss, it is included using a simplified procedure. As we show in \cite{Smolec-2026} (see Sect.~3.5 and Fig.~5 therein) evolutionary tracks computed with the Reimers mass-loss prescription are only weakly sensitive to the adopted efficiency parameter. In particular, mass loss has a negligible effect on the luminosity level of the blue loops; the differences between tracks computed with and without mass loss are much smaller than, for example, the luminosity uncertainties of our systems. The primary effect of mass loss is a reduction of the stellar mass itself. However, the expected amount of mass lost is also small, being comparable to the uncertainty in the mass determination. Consequently, for the final model grid of each star, we apply a constant correction to the initial ZAMS mass based on the expected evolutionary stage inferred from the initial model grid. The subsequent evolution is then computed including mass loss with $\eta=0.4$. Further details and justification of this procedure are provided in Appendix~\ref{secapp:massloss}. Owing to the negligible impact of mass loss on the evolutionary tracks in the HR diagram, its inclusion has virtually no effect on the solutions selected for the systems discussed below.

Finally, for the initial model grid, solutions are selected based on \chis{} including \Teff{}, $L$, and $R$, but not the pulsation period, as was done for these systems in previous work. However, we check whether for the identified solutions the pulsation period also matches the observed value. In the final model grid, we used the pulsation period to constrain the models. In the following section, we discuss the determination and precision of the computed pulsation periods.

\subsection{Determination of pulsation periods}\label{ssec:puls}

There are two types of periods we can obtain from \RSP: linear and nonlinear. The first is based on linear non-adiabatic (LNA) perturbation analysis of the static envelope. Nonlinear period corresponds to full amplitude regime and requires direct time integration of the model. It is the nonlinear pulsation period that should be compared with observations, but its computation is time consuming, even for a single model. Fortunately, for classical Cepheids, the expected difference is small, below 1\%, as we demonstrate later; see also \cite{Bono-1999}. Nonlinear correction depends mostly on physical parameters and amplitude of pulsation, and thus it is a slowly varying function of the position in the HRD. As we needed to the compute pulsation periods for tens of models along evolutionary tracks, we used linear periods in our modeling with a constant nonlinear period correction applied. In this section we calculate the nonlinear period correction for each of the Cepheids and examine how precise the computed linear periods are. 

We defined the reference pulsation model as one computed using the default setup described in Sect.~\ref{ssec:evpuls}. To calculate the nonlinear period for a given Cepheid, we adopted the reference pulsation setup and set stellar parameters ($M$, \Teff{}, and $L$) to the central values from Tab.~\ref{tab:dlebs}, with $Z$=0.006. Linear periods, $P_0$, derived from LNA analysis of the static envelope are collected in the third column of Tab.~\ref{tab:P}. Models are then integrated until full amplitude is reached and the period stabilizes, and subsequently for additional cycles to ensure convergence. The period typically stabilizes after about 1000 cycles, and models are evolved for $\sim$6000 cycles in total. The resulting nonlinear periods, $P_{\rm nl}$, are collected in the fourth column in Tab.~\ref{tab:P}, followed by nonlinear period correction, $\delta P=(P_{\rm nl}-P_0)/P_0$. For CEP-0227 and CEP-4506 the correction is of similar order, 0.28\% and 0.21\%, respectively. For CEP-1812, the least luminous F-mode Cepheid in the sample, most likely on the first crossing, the correction is negligible. For the 1O Cepheids the corrections are $\leq 0.1\%$. 

\begin{table}
\caption{Pulsation period and radius values from reference static, linear and nonlinear, and full-amplitude pulsation models. }
\label{tab:P}
\centering
\small
\setlength{\tabcolsep}{4pt}
\begin{tabular}{lrrrrrr}
\hline\hline
ID & Mode & $P_0$ & $P_{\rm nl}$ & $\delta P$ & $R_0$ & $R_{\rm nl}$ \\
   &      & (d)   & (d)          & (\%)       & ($R_\odot$) & ($R_\odot$) \\
\hline
CEP-0227         & F    & 4.0386 & 4.0498  & 0.28 & 35.00 & 35.38 \\
CEP-4506         & F    & 3.1444 & 3.1509  & 0.21 & 28.58 & 28.93 \\
CEP-1812         & F    & 1.4244 & 1.4244  & 0.00 & 18.00 & 18.18 \\
CEP-2532         & 1O   & 2.1985 & 2.2007  & 0.10 & 29.45 & 29.51 \\
CEP-1718A        & 1O   & 1.9177 & 1.9184  & 0.04 & 27.81 & 27.92 \\
CEP-1718B        & 1O   & 2.5800 & 2.5815  & 0.06 & 33.12 & 33.20 \\
\hline
\end{tabular}
\tablefoot{The columns present the Cepheid's ID; mode of pulsation; linear and nonlinear period, $P_0$ and $P_{\rm nl}$; nonlinear period correction, $\delta P$; and static and nonlinear radius, $R_0$ and $R_{\rm nl}$.}
\end{table}

Our modeling strategy assumes computing linear pulsation periods for tens of models along evolutionary tracks, and applying a nonlinear period correction before comparing the period with the observed one. This is the only feasible procedure to include pulsation period in the modeling, as computation of nonlinear models to full amplitude is extremely time consuming.

\begin{figure*}
    \centering
    \includegraphics[width=\linewidth]{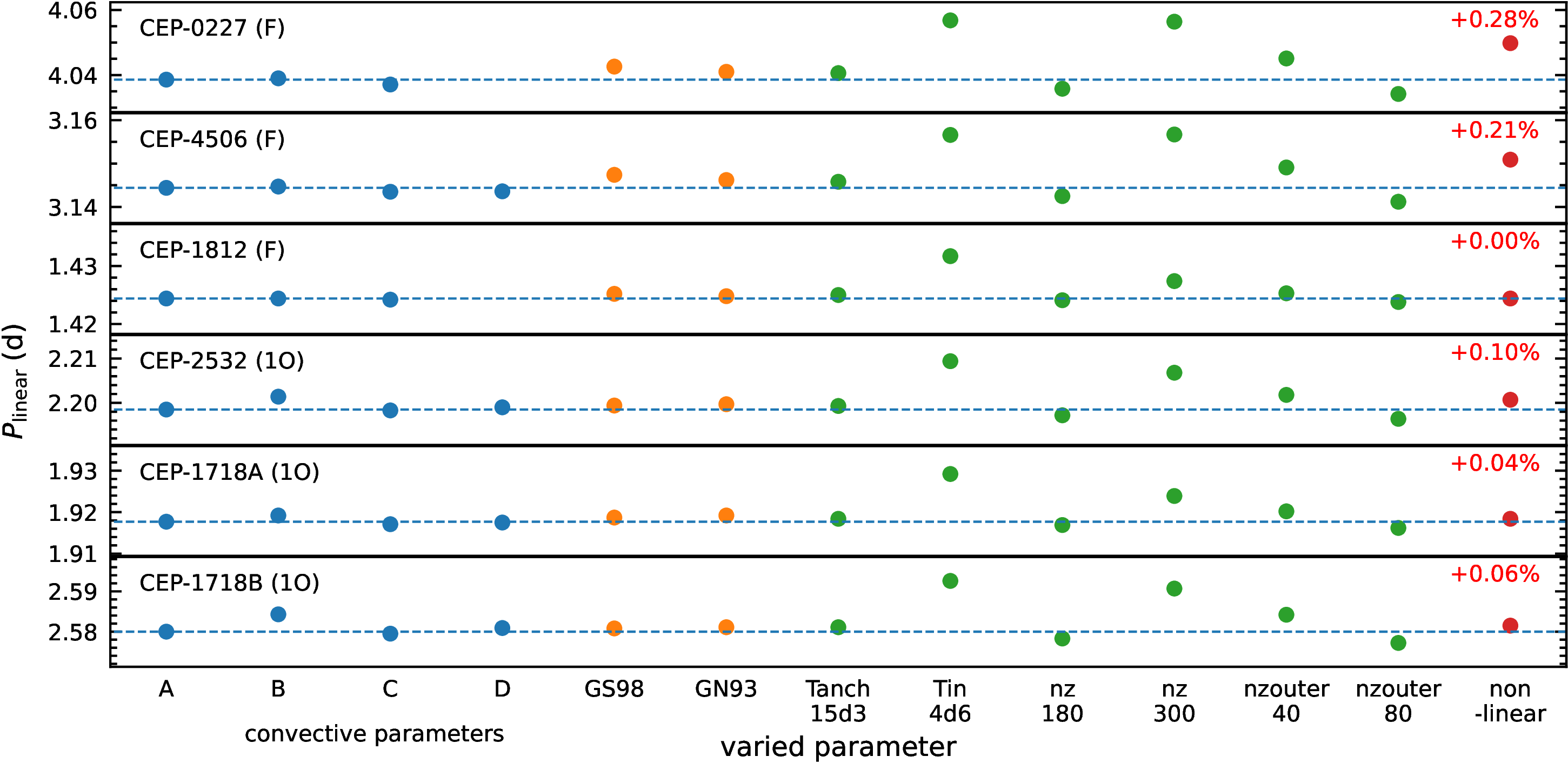}
    \caption{Linear pulsation periods computed by varying different parameters in the reference model (first point and horizontal dashed lines): convective parameter set \citep[A--D; see tab.~4 in][]{Paxton-2019}, solar metal mixture (\citet{GS-98}, GS98, \citet{GN-93}, GN93), and envelope grid parameters (\texttt{T\_anch}, \texttt{T\_in}, \texttt{nz}, and \texttt{nzouter} see Sect.~\ref{ssec:evpuls}). The last points represent the nonlinear periods assuming the reference settings. The nonlinear period correction is given above the last point.}
    \label{fig:linpuls}
\end{figure*}

We now check how precise linear pulsation periods are. To this end, we run linear models again, varying one parameter at a time from the reference model. The results are presented in Fig.~\ref{fig:linpuls}, with one panel for each star, the period on the y-axis and the model label on the x-axis. In each panel, the first point corresponds to the reference model, subsequent points correspond to the varied models, and the last (red) point corresponds to the nonlinear model. Among the parameters that we varied are the convective parameter set \citep[sets A--D from Tab.~4 in][]{Paxton-2019}, the reference solar mixture, and the parameters of the envelope grid. The linear pulsation periods are most sensitive to changes in the grid parameters. Still, the largest relative change in the pulsation period, $\Delta P/P_{\rm ref}$, where $P_{\rm ref}$ is the period of the reference model, is only 0.6\%.

Accuracy of the computed linear periods is another issue, and a more difficult one to address. A comparison with an independent, potentially more accurate code, such as \gyre{} \citep{Townsend-2013}, would be useful; however, it cannot be performed in a strictly one-to-one manner. While \gyre{} uses the full stellar structure from evolutionary calculations, its treatment of convection is simplified (frozen-in approximation, with convective stratification taken from MLT). \RSP{}, on the other hand, is an envelope code, but includes pulsation–convection coupling. For a few evolutionary tracks, selected to match the positions of the considered Cepheids, we computed pulsation periods with \gyre{} and then with \RSP{}, adopting the same global parameters. Periods from \gyre{} are shorter by 0.7–1.4\%. Taking into account the intrinsic differences between the two approaches, we conclude that the agreement is very satisfactory.

In the following we assume the periods are calculated with 1\% uncertainty. Later in Sect.~\ref{ssec:pvsr} (see also Appendix~\ref{secapp:RobustTension}), we also investigate how sensitive our conclusions are to this assumption as well as to the possible systematic shift in the computed pulsation periods.

\subsection{Selection of best-matching models}\label{ssec:chisq}

A common method to select the best-matching evolutionary model employs either Bayesian approach or \chis{} minimization in the plane of the HRD \citep{Deka-2025, Cassisi-2011, Neilson-2012, Prada-Moroni-2012}. We used \chis{} minimization. The \chis{} was computed separately for the two components of the system using
$$ \chis = \sum_{i=1}^{n} \left( \frac{O_i-E_i}{\sigma_i} \right)^2, $$
where $O_i$ are the observed parameters, $\sigma_i$ are the associated errors, and $E_i$ are the model predictions. In \cite{Deka-2025}, \Teff{}, $L$, and $R$ were included in \chis{}. We used the same approach in Sect.~\ref{ssec:gridone}, but in Sect.~\ref{ssec:gridtwo}, we also include the pulsation period. 

The three parameters $T_{\rm  eff}$, $L$, and $R$ are not independent. Since $L\propto R^2 T_{\rm eff}^4$, the use of all three in \chis{}  may seem redundant. However, from the observational point of view, these quantities are derived from different observables and are determined with different precision. Thus, including all of them in \chis{} represents not only a consistency test for the observables, but also utilizes all available observational information about the system.

A similar argument applies to the pulsation period. Because of the period–mean density relation, it provides information similar to that of the stellar radius. However, the determination of the pulsation period is completely independent of the radius determination, and its precision is orders of magnitude higher. So far, it has been rarely used in modeling of Cepheids in binary systems; a recent exception is \cite{Espinoza-2025}, where a method combining evolutionary and pulsation calculations was proposed to constrain parameters of components of binary system, including double-mode Cepheid, for which only mass ratio was determined from observations.

In the actual calculation of \chis, for $\Teff$, $\log L$, $R$, and period, $P$, we use the data from Tab.~\ref{tab:dlebs}. The corresponding model parameters, $E_i$, are derived from the evolutionary tracks. Linear pulsation periods are computed with \RSP{} along the evolutionary tracks, and a nonlinear period correction is applied prior to comparison with the observed values. From the observational point of view, the pulsation period is extremely precise \citep[uncertainties are of order of $10^{-6}$d for typical OGLE data; see, e.g.,][]{Smolec-2023}; in \chis{} we adopt a 1\% uncertainty, which reflects the theoretical uncertainty of its determination (see Sect.~\ref{ssec:puls}). The model parameters are interpolated along the tracks at intervals of 0.001\,Myr. For non-pulsating stars we interpolate along the entire track, while for Cepheids we restrict the interpolation to the immediate vicinity of the IS.

As mentioned above, the parameters we include in $\chi^2$ are not independent, and the covariance matrix is unavailable. In addition, for some of the systems we used additional constraints, for example, requiring one of the components to be more evolved, in agreement with the system mass ratio, or located on a specific crossing. Constraints are also applied to the theoretical models; by default, we require the metallicity and overshooting parameters to be the same for both components of a given system (but not necessarily the same as for other systems considered). Consequently, $\chis$ does not have its usual statistical meaning and is used only to identify the best-matching models, to which the additional constraints may be applied.

We expect that the two components of the system have the same age. The masses, although determined quite precisely, are uncertain by about 0.02 to 0.1\MS{} (see Tab.~\ref{tab:dlebs}). In the modeling, when matching the two stars, we therefore allowed for a small age difference to account for these uncertainties. The MS evolutionary timescale scales approximately as $\tau\propto{}M^{-2.5}$ \citep[e.g.,][]{SSE}, which implies $\Delta\tau/\tau$$\approx$2.5$\Delta M/M$. This relation can be used to estimate the allowable age difference arising from uncertainties in the component masses. We adopt a conservative estimate, $\Delta M$=2$\sigma_M$. For the evolutionary timescale, we adopt $\tau$$\approx$160\,Myr, which is approximately the MS lifetime for the mass/metallicity range considered \citep[4\MS, $Z$=0.006; see Tab.~6 in][]{Smolec-2026}. Although our systems are evolved, most often in the core helium-burning phase, the spread in their evolutionary ages is largely determined by differences in their MS lifetimes, making this a reasonable first-order approximation. We note that the similar estimates were adopted by \cite{Neilson-2012}. Since the masses of our components are very similar, and the uncertainties for both components are either the same or differ by only 0.01\MS, the result depends very little on which mass/uncertainty is used in the calculation for a given system (differences are below 1\,Myr). With this procedure, the maximum allowed age differences are very similar for CEP-0227 and CEP-4506; in the following we adopt 6\,Myr. For CEP-2532 we obtain $\approx$8\,Myr. Due to the largest mass uncertainty, the allowed age difference is also the largest for CEP-1718 and amounts to $\approx$20\,Myr. 

We stress that the above procedure is highly simplified, and the resulting estimates should not be regarded as robust constraints on the models. Rather, they provide an approximate benchmark for assessing whether the inferred age differences are consistent with the observational uncertainties in the component masses.

\section{Results}\label{sec:results}

\subsection{Solutions for initial model grid}\label{ssec:gridone}

\begin{figure*}
\centering
\includegraphics[width=\linewidth]{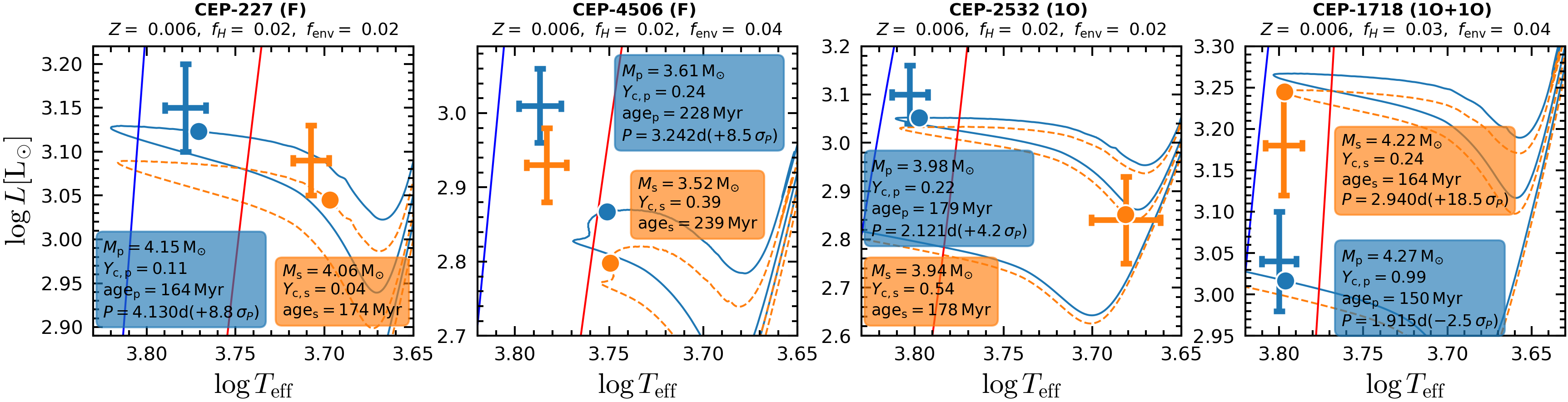}
\caption{Solutions for initial model grid with a fixed $Z$=0.006 and common overshooting parameters for the system's components. The name of the system and the values of $Z$, \fcor{}, and \fenv{} are given above each panel. Primary (more massive) component is marked in blue, the secondary in orange. Crosses correspond to observations, and filled circles correspond to the best-matching solution. In the colored boxes some of the best-matching solution's parameters are given: central helium content ($Y_{\rm c}$) and age. In the case of Cepheids, the period and its offset from the observed value are also given.}
\label{fig:grid1}
\end{figure*}

\begin{figure*}
\centering
\includegraphics[width=\linewidth]{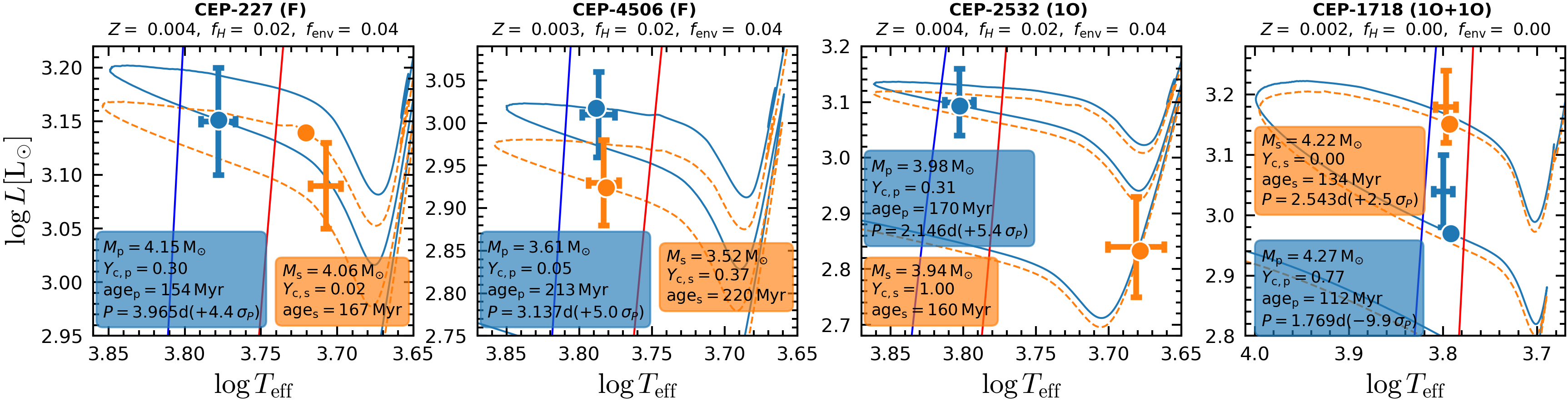}
\caption{Same as Fig.~\ref{fig:grid1} but with the constraint on metallicity lifted.}
\label{fig:grid1B}
\end{figure*}

In this section, best-matching models are selected based on \chis{} including $\log L$, \Teff{}, and $R$. At this stage, additional observational constraints (e.g., PCR) are not imposed on the solution but are used for their qualitative interpretation. For Cepheids, we also record pulsation period, corrected by nonlinear period correction (from Tab.~\ref{tab:P}). We note how far the model period is from the observed value, in terms of 1\% uncertainty of model period (Sect.~\ref{ssec:puls}). This information is given in Figs.~\ref{fig:grid1} and \ref{fig:grid1B} and often points to significant discrepancies, as we discuss at the end of this section and later in Sect.~\ref{ssec:pvsr}.

We first examine whether satisfactory solutions can be found assuming average metallicity of the LMC, $Z$=0.006 (Sect.~\ref{ssec:data}). We show the best solutions in Fig.~\ref{fig:grid1}, each system on a separate panel. 

In the solution for CEP-0227, both components are placed on the upper branch of the loop. The Cepheid is on the third crossing, and the companion is already beyond the red edge. The less massive companion is more advanced, and it is at the very end of the CHeB. This is in conflict with basic stellar physics, as the more massive star, the Cepheid, should be more evolved. Assuming $Z$=0.006, we can find a solution in which the companion is less advanced. However, this physically justified solution is significantly worse in terms of matching the positions of the components on the HRD. Moreover, for these solutions the age difference is more than twice the estimated maximum allowed value of 6\,Myr. We conclude that in the initial grid there is no satisfactory solution for CEP-0227, assuming the mean LMC metallicity.

For CEP-4506, with $Z$=0.006, the blue loops are too dim and not extended enough to reach the observed error bars. Consequently, no satisfactory solutions can be found.

The solution for CEP-2532 at $Z$=0.006 is satisfactory. The Cepheid is located on the third crossing, close to the maximum extent of the blue loop, while the companion lies on the lower branch of the blue loop, already burning helium in the core but still relatively close to the RGB. The age difference is only 1\,Myr, and the more massive star is more evolved.

As mentioned in Sect.~\ref{ssec:data}, CEP-1718 is problematic to model because the more massive star is less luminous. The best solution found at $Z$=0.006 places the more massive primary on the first crossing and the secondary on the blue loop. Moreover, solutions on the first crossing are less probable due to the expected rapid evolutionary timescale, which should have already been apparent in the OGLE data. We also note that this first crossing solution was facilitated by very high core overshooting, \fcor=0.03. Overall, we do not find a satisfactory solution for CEP-1718 at $Z$=0.006.

Since, for some systems, we could not obtain satisfactory solutions when assuming the average LMC metallicity, we now relax this constraint, still requiring the same $Z$ for both stars. This variant of the solutions is presented in Fig.~\ref{fig:grid1B}.

After relaxing the metallicity constraint, the best solution for CEP-0227 now places the Cepheid on the second crossing, while the companion remains on the upper branch of the blue loop. The position of the Cepheid on the HRD is now matched almost perfectly with $Z$=0.004. The less massive star remains evolutionarily more advanced, and the age difference is even larger than in the previous $Z$=0.006 solution.

For CEP-4506, a much lower metallicity, $Z$=0.003, is required to produce longer loops that can now cross the IS and match the positions of both components. The Cepheid is found on the third crossing, with core helium nearly depleted, while the non-pulsating, less massive companion is on the second crossing. The age difference is 7\,Myr, approximately equal to the estimated limit. The solution is therefore satisfactory.

In the case of CEP-2532, the best overall solution is obtained for a lower metallicity of $Z$=0.004. The Cepheid is again on the blue loop, but the companion is located on the RGB. As a result, the age difference is significantly larger, reaching 10\,Myr, slightly exceeding the estimated limit of 8\,Myr.

For CEP-1718, tracks with the lowest metallicity in the grid, $Z$=0.002, provide the best match to the positions on HRD. At this metallicity, the loops are wide and extended, allowing for a good fit to both components. The primary is found on the second crossing, and the secondary on the third crossing. However, an unavoidable issue with this system is that the more massive star is less luminous. Moreover, the age difference, 22\,Myr, slightly exceeds the estimated limit of 20\,Myr. Finally, the PCR of the secondary places it on the second crossing rather than the third (although this is not a strong constraint; see Sect.~\ref{ssec:data}).

When we allow for metallicities different from the adopted mean LMC value, $Z=0.006$, the best-fitting solutions for all systems indicate lower metallicity. This does not necessarily imply that the actual metallicity is indeed lower. Rather, it may result from the degeneracy between metallicity and mixing processes, or more generally reflect the mass discrepancy problem. We return to this issue in Sect.~\ref{ssec:gridtwo}.

We note that, for most of the solutions presented in this section, the best-matching models adopt the same efficiency of envelope overshooting, \fenv=0.04. The exceptions are the solutions for CEP-0227 and CEP-2532 in Fig.~\ref{fig:grid1}, for which \fenv=0.02, and the solution for CEP-1718 in Fig.~\ref{fig:grid1B}, for which \fenv=0.0. However, for both CEP-0227 and CEP-2532, solutions with $Z$=0.006 and \fenv=0.04 are only slightly worse than those displayed in Fig.~\ref{fig:grid1}. At low metallicity (the case of CEP-1718, with $Z$=0.002), the solutions depend very weakly on \fenv, and marginally worse (in terms of \chis{}) solutions can be found for any considered value of $\fenv$.

\begin{figure}
\centering
\includegraphics[width=\linewidth]{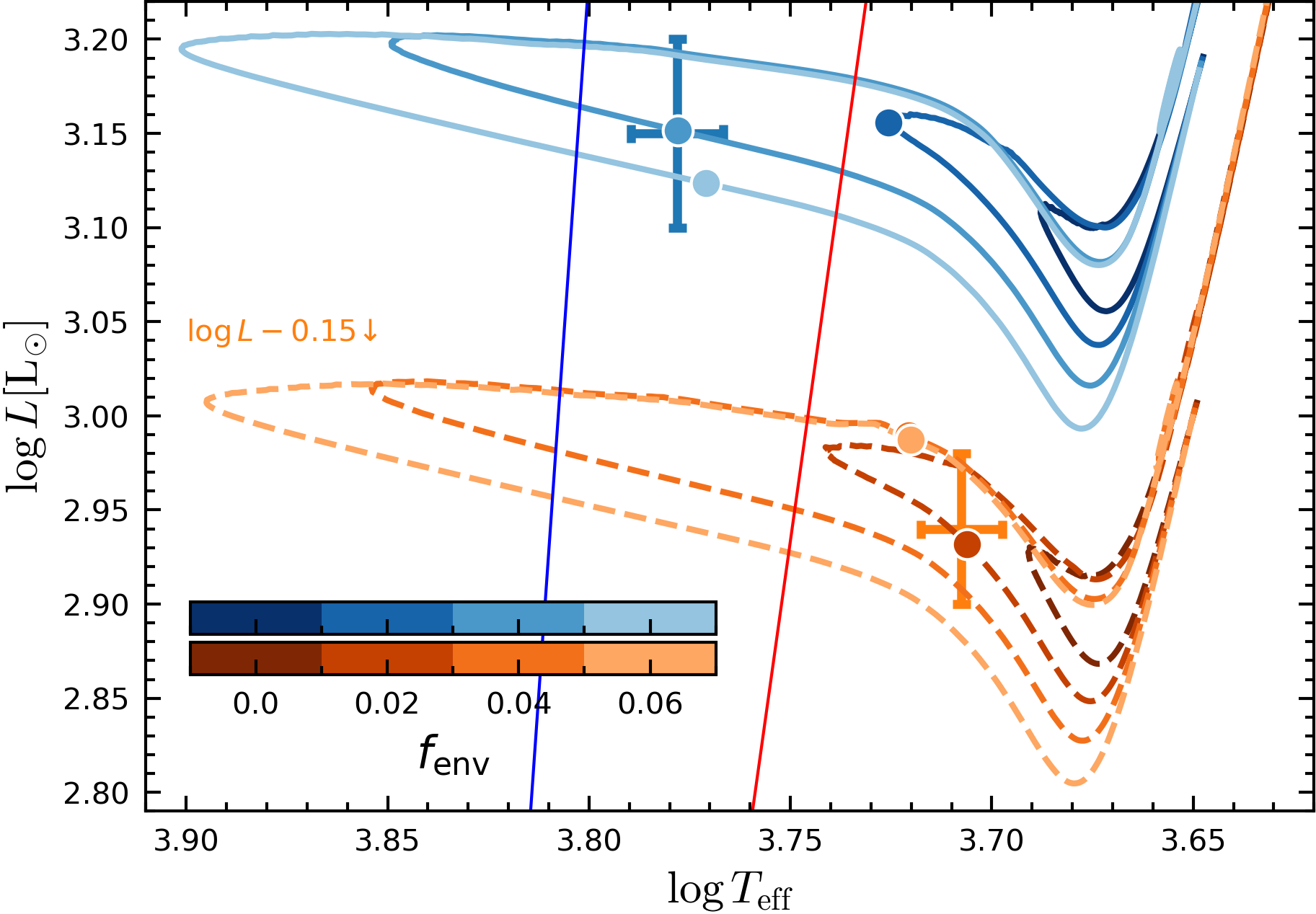}
\caption{Effects of changing \fenv{} on evolutionary tracks with $Z$=0.004 and \fcor=0.02, aimed at matching CEP-0227 (error bars). Cepheid tracks are shown in shades of blue, and companion tracks (shifted by $-0.15$ in $\log L$, for clarity) are in shades of orange. The colors correspond to \fenv. Filled circles mark the best solutions (for the companion, solutions for \fenv=0.04 and \fenv=0.06 overlap).}
\label{fig:fenv}
\end{figure}

We note that envelope overshooting acts only during RGB evolution. The effects on pulsation during the crossing of the IS are indirect, through the modified shape of the evolutionary tracks. In general, higher $\fenv$ leads to longer loops, especially for high-metallicity tracks (see \citealt{Smolec-2026}, for a detailed analysis, in particular their Sect.~3.4 and Fig.~4). We also illustrate the effects of varying $\fenv$ in Fig.~\ref{fig:fenv} for CEP-0227. The tracks shown in the figure all have $Z$=0.004 and \fcor=0.02, but differ in \fenv=(0.0, 0.02, 0.04, 0.06). The corresponding best-matching solutions are marked with filled circles. For low values of $\fenv$ (\fenv$\leq$0.02), the loops are too short to reach the IS. On the other hand, for \fenv$\geq$0.04, the solutions we obtain are very similar; for the secondary, they effectively overlap in the figure.

For these reasons, envelope overshooting is fixed at \fenv=0.04 in the final model grid, as this value ensures that the loops are sufficiently extended to reach the IS. As demonstrated by \cite{Smolec-2026}, this value provides reasonable loop extents over a wide range of masses, metallicities, and MS core overshooting efficiencies. For $\fenv\geq0.04$, the resulting solutions and their \chis{} values depend only weakly on the adopted value of \fenv.

We note that for nearly all solutions presented in this section, the efficiency of MS core overshooting is at least \fcor=0.02. The only exception is the solution for CEP-1718 shown in Fig.~\ref{fig:grid1B} with \fcor=0.0; however, as discussed above, this solution is unrealistic. Furthermore, when metallicity is allowed to vary from the average LMC value, all resulting solutions favor lower metallicities. The effects of changing $Z$ and $\fcor$ on the luminosity of the blue loops are degenerate: both lower $Z$ and higher $\fcor$ lead to increased luminosity (see Sect.~\ref{ssec:data}). A more general discussion of these trends is presented after analysis of the solutions obtained with the final model grid.

For all systems, ``good'' solutions can be found that reproduce the observed positions on the HRD and the radii of both components, either with fixed $Z$=0.006 (Fig.~\ref{fig:grid1}) or with metallicity treated as a free parameter (Fig.~\ref{fig:grid1B}). However, only for CEP-2532 and CEP-4506 (in particular for $Z$=0.004) do these solutions satisfy all observational constraints, including the luminosities, the PCR (for CEP-1718), and a consistent evolutionary ordering of the components.

We therefore attempted to find improved solutions using final model grids tailored to each system and based on the initial solutions presented above. These grids are more localized and denser in both $Z$ and $\fcor$. For CEP-0227, CEP-4506, and CEP-2532, the central metallicity is the same as in Fig.~\ref{fig:grid1B}, and six additional values are included with a spacing of $dZ$=0.0005. For \fcor, we consider three values, \fcor=(0.016, 0.02, 0.024), while \fenv=0.04. For the most challenging system, CEP-1718, we explore a significantly broader range in both metallicity ($Z$=0.001–0.007, with a step $dZ$=0.001) and $\fcor$ (\fcor=0.010, 0.015, 0.020, 0.025).

For all systems, we explore masses slightly offset from the central values. For each component, we adopt five values: $M_0$, $M_0 \pm 0.5\sigma_M$, and $M_0 \pm \sigma_M$. Mass loss on the RGB is also included, with $\eta$=0.4 and mass correction applied at ZAMS (Appendix~\ref{secapp:massloss}).

Before analyzing the final grid, we note a serious issue present in all solutions discussed so far. In Figs~\ref{fig:grid1} and \ref{fig:grid1B}, for each modeled Cepheid we provide the pulsation period for the best-matching solution and quantify its deviation from the observed value in units of the assumed model uncertainty of 1\% (Sect.~\ref{ssec:puls}). Focusing on solutions that reproduce the Cepheid positions on the HRD, we find a systematic and significant discrepancy: the predicted periods are consistently too long -- by about 4 to $8\sigma$ (CEP-0227), $5\sigma$ (CEP-4506, CEP-2532), and $3\sigma$ (CEP-1718B). The only exception is CEP-1718A, for which the period is significantly shorter ($-3\sigma$, $-10\sigma$). Attempting to include the pulsation period in the $\chis$ alongside the radius introduces a clear tension between the two constraints. No satisfactory solution that simultaneously matches both the pulsation period and the radius can be found. This issue is investigated in the following section.

\subsection{The tension between pulsation period and radius on HRD}\label{ssec:pvsr}

\begin{figure*}
\centering
\includegraphics[width=.5\linewidth]{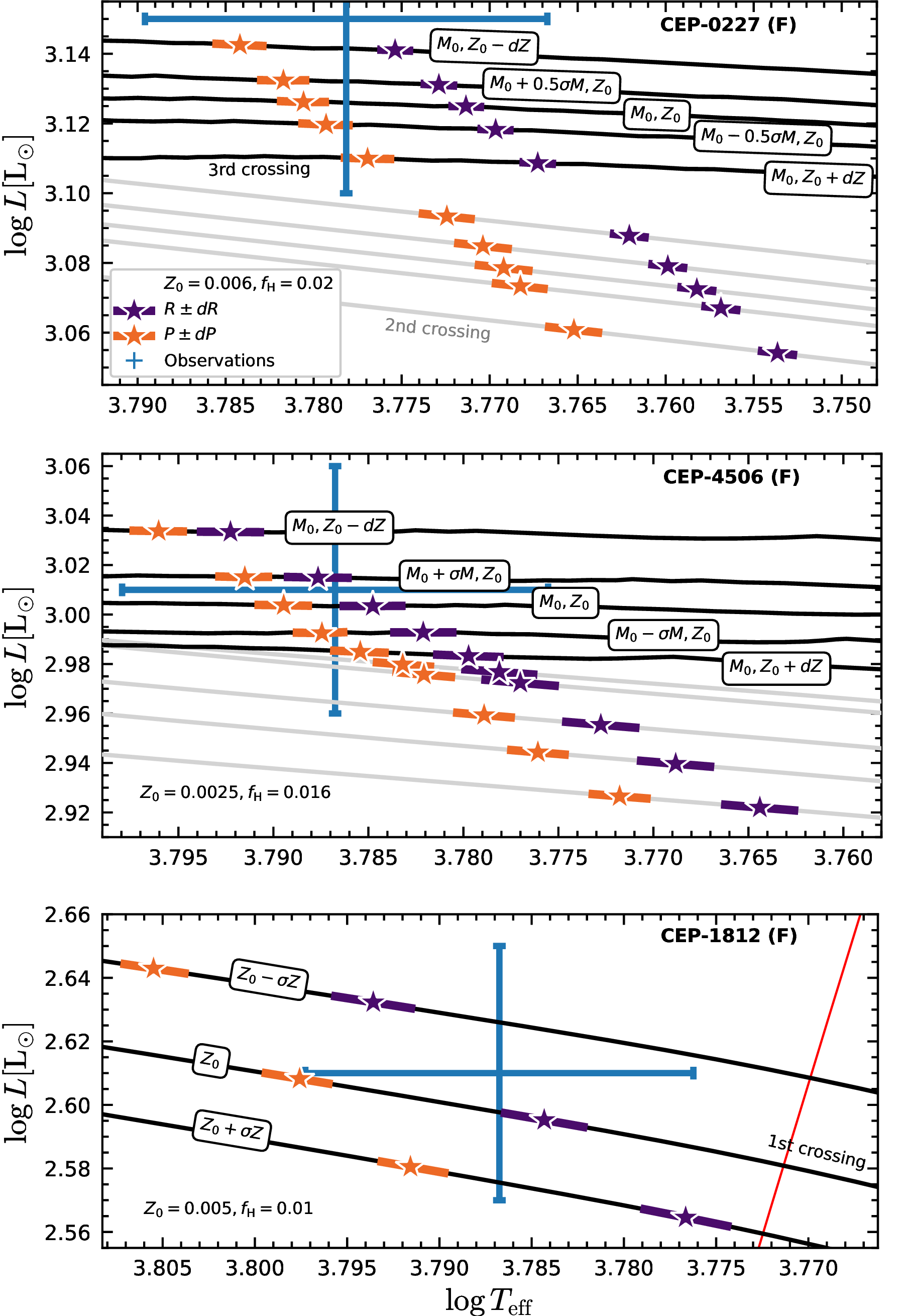}\includegraphics[width=.5\linewidth]{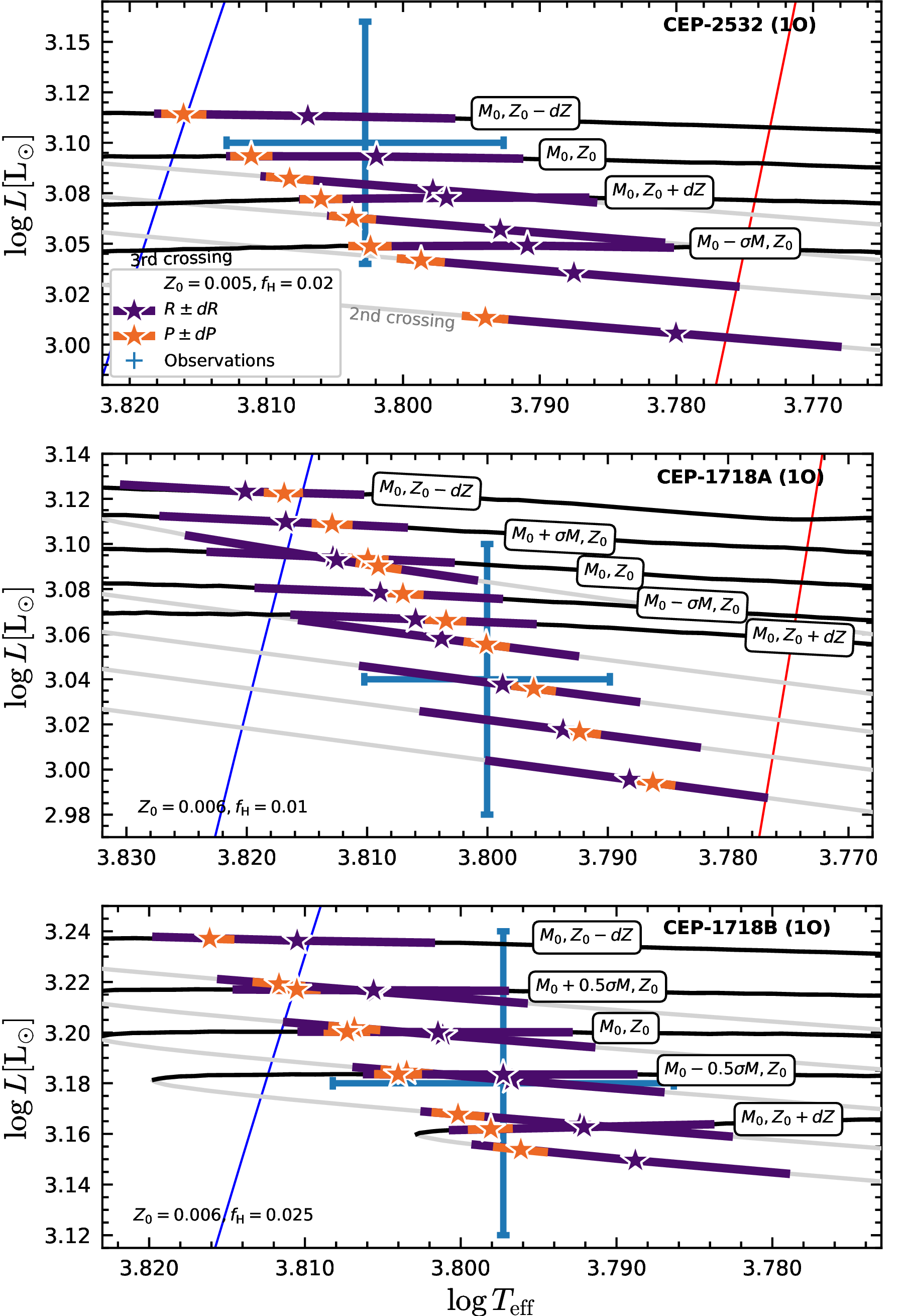}
\caption{Loci where the pulsation period is matched within 1\% (orange asterisk and thick line segment) and where the stellar radius is consistent within its observational uncertainty (purple asterisk and thick line segment), shown along evolutionary tracks in the vicinity of the observed position of the F-mode (left) and 1O Cepheids (right) on the HRD (error box).}
\label{fig:PRtensionFO}
\end{figure*}

The tension between pulsation period and radius on HRD is investigated in Fig.~\ref{fig:PRtensionFO}, for F-mode and 1O Cepheids (left and right panels, respectively). Using evolutionary tracks from the final grid, we select a family centered on a track with central mass $M_0$ and such $Z$ and $\fcor$ that it crosses the observational error box of the Cepheid in the HRD. Additional tracks have slightly different masses and metallicities, $\pm dZ$ and $\pm\sigma_M$. The results are not sensitive to the exact values of $Z$ and $\fcor$. We then interpolate the location where the observed pulsation period is matched (orange asterisk; nonlinear period correction is taken into account) and mark the 1\% range (thick orange segment). Similarly, we interpolate the location matching the observed radius (purple asterisk) and mark the $1\sigma$ range (thick purple segment).

For F-mode Cepheids, the tension between $P$ and $R$ is immediately apparent. When we match the pulsation period, the radius is too small; when we match the radius, the pulsation period is too long. For CEP-0227, at the location where pulsation period is matched, the radii are off, on average by $-12.0\sigma_R$ (where $\sigma_R$ is the observational error from Tab.~\ref{tab:dlebs}). For CEP-4506 the corresponding number is $-3.2\sigma_R$. It is also clear that this tension cannot be easily removed by adjusting metallicity, mass, or the efficiency of overshooting. The locations at which the radius and pulsation period are matched trace two nearly parallel lines, as expected. The position of the constant-radius line follows from $R\propto L^{1/2}T_{\rm eff}^{-2}$, while, due to the period-mean density relation, the constant-period line is roughly parallel to it. Unfortunately, the two do not coincide.

To increase the sample, in the left panels of Fig.~\ref{fig:PRtensionFO}, alongside CEP-0227 and CEP-4506, we include the F-mode Cepheid in CEP-1812. We do not model CEP-1812 as a system, due to binary interaction (see Sect.~\ref{sec:intro}); however, the Cepheid itself, with a precisely determined radius, provides an additional test of the discrepancy. We find the same systematic offset,$-6.2\sigma_R$ on average, at the location where pulsation period is matched.

While the systematic discrepancy is clear, we note that points matching the pulsation period, and matching the radius, can still be found within the \Teff{}/$L$ error box in the HRD. Good agreement can be achieved for $L$, $\Teff$, and $P$, and separately for $L$, $\Teff$, and $R$, but not for $R$ and $P$ simultaneously. This reflects the fact that $R$ and $P$ are the most precisely determined quantities. For F-mode Cepheids, $R$ is measured  with a typical precision of 0.7\% or better (Tab.~\ref{tab:dlebs}). Pulsation periods from OGLE data are typically accurate to $\sim$10$^{-6}$\,d or better. For $P$, however, the dominant uncertainty is theoretical (1\%, Sect.~\ref{ssec:puls}).

For 1O Cepheids (right panels in Fig.~\ref{fig:PRtensionFO}), the situation is slightly different, mainly due to the significantly lower precision of their radii, which are an order of magnitude less accurate (see Tab.~\ref{tab:dlebs}). As a consequence, the portions of the evolutionary tracks that match the radius within its uncertainty are extended, covering an effective temperature range comparable to the uncertainty in \Teff{} determination. This strongly reduces the apparent tension. Still, for CEP-2532 and CEP-1718B, when matching the pulsation period, the radius is systematically too small (by $-0.9\sigma_R$ and $-0.7\sigma_R$, respectively). CEP-1718A is the only case where the two quantities well agree; in this case, the radius is slightly larger (+0.3$\sigma_R$) when the pulsation period is matched.

The 1O sample does not contradict the existence of a tension between solutions based on the radius and those based on the period; rather, the observational uncertainties are simply too large to reveal it clearly. For CEP-2532 and CEP-1718B, keeping the central value of the radius but reducing its uncertainty would lead to a picture similar to that observed for F-mode Cepheids. More precise radius determinations are therefore needed for 1O stars to assess whether, and to what extent, this tension is present.

The reported tension is robust, that is, it cannot be removed by assuming a larger uncertainty in the determination of pulsation periods or by introducing a systematic shift in their values. We demonstrate this in Appendix~\ref{secapp:RobustTension}, where Fig.~\ref{fig:PRtension_mod} presents a modified version of Fig.~\ref{fig:PRtensionFO}, assuming a 1.5\% uncertainty in the pulsation periods (nearly three times the value estimated in Sect.~\ref{ssec:puls}) and pulsation periods systematically shorter by 1.4\% at a given position in the HR diagram. This value corresponds to the largest difference between the linear pulsation periods computed with \RSP{} and \gyre{} (see Sect.~\ref{ssec:puls}). For this direction of the systematic shift, the discrepancy is indeed slightly reduced, but remains large and significant. For the F-mode Cepheids, the tension decreases from $-12.0\sigma_R$ to $-9.9\sigma_R$ (CEP-0227), from $-3.2\sigma_R$ to $-2.1\sigma_R$ (CEP-4506), and from $-6.2\sigma_R$ to $-5.2\sigma_R$ (CEP-1812). For the 1O Cepheids, the corresponding changes are from $-0.9\sigma_R$ to $-0.8\sigma_R$ (CEP-2532), from $0.3\sigma_R$ to $0.5\sigma_R$ (CEP-1718A), and from $-0.7\sigma_R$ to $-0.5\sigma_R$ (CEP-1718B).

What is the origin of the discrepancy between solutions based on matching the stellar radius and the pulsation period? For the comparison between observed and model values to be meaningful, it is essential that the same quantities are being compared. This condition is well satisfied for the pulsation period, which is a well-defined quantity both observationally and theoretically. The only subtlety is that the observed period corresponds to the nonlinear, full-amplitude value, rather than to the linear period derived from a stability analysis of a static envelope. However, the appropriate correction has been applied prior to the comparison with the models.

In the case of the radius, the observed value has so far been directly compared with the radius along the evolutionary track, and we argue that this is one of the sources of the discrepancy. The radius inferred from evolutionary models corresponds to that of a static star. As in the case of the period, the radius corresponding to a pulsating model at full amplitude may differ. To our knowledge, this issue has not been addressed in the literature so far, with the exception of analysis based on amplitude equations, which explicitly predicts a nonlinear modification of the radius \citep[][]{BG-1984}.

In the last two columns of Tab.~\ref{tab:P}, we list the radius corresponding to the static model for each Cepheid as well as the radius from the nonlinear model -- the same \RSP{} model used to determine the nonlinear period correction. The radius in the nonlinear case was derived as the zeroth-order term of a Fourier series fitted to the temporal variation of the model radius. In all cases, it is larger than the static radius. The difference amounts to 0.38\RS{} for CEP-0227, corresponding to 3.2 times the observational uncertainty of its radius (3.2$\sigma_R$). The corresponding values for CEP-4506 and CEP-1812 are 0.35\RS{} (1.8$\sigma_R$) and 0.18\RS{} (1.4$\sigma_R$), respectively.

For 1O Cepheids -- CEP-2532, CEP-1718A, and CEP-1718B -- the differences are significantly smaller: 0.06\RS{} (0.04$\sigma_R$), 0.11\RS{} (0.09$\sigma_R$), and 0.08\RS{} (0.06$\sigma_R$), respectively. The very small differences in units of $\sigma_R$ are a consequence of the roughly ten times larger observational errors in the 1O sample. The results, however, clearly indicate that the correction depends on the pulsation mode.

The above results suggest that at least part of the tension arises from an inconsistent comparison between model and observed radii. Assuming that we correctly match the observed and model periods, the tension is in fact, between the theoretical and empirical radii. 

For F-mode Cepheids, if a nonlinear correction to the radius is applied at the point along the evolutionary track where the pulsation period is matched, the tension is reduced to (average for points displayed in Fig.~\ref{fig:PRtensionFO}) 8.8$\sigma_R$ for CEP-0227, 1.4$\sigma_R$ for CEP-4506, and 4.8$\sigma_R$ for CEP-1812. The numbers for 1O Cepheids are barely affected, but again it is a consequence of an order of magnitude lower precision of radii determination.

The nonlinear radii given above were computed assuming the convective parameters of set A. For all considered Cepheids, we also computed nonlinear pulsation models assuming the convective parameter sets B, C, and D. The resulting increases in radius relative to the static value are comparable. For CEP-0227, for example, the increase varies from 0.9\% (2.8$\sigma_R$, set C) to 1.1\% (3.2$\sigma_R$, sets A and D). A larger spread is found for CEP-4506, for which the increase varies from 0.8\% (1.2$\sigma_R$, set C), through 1.2\% (1.8$\sigma_R$, set A; the value adopted above), to 1.5\% (2.2$\sigma_R$, set B). In this case, adopting the largest value reduces the tension from 1.4$\sigma_R$ to 1.0$\sigma_R$. For CEP-1812, the value obtained for set A, used above, is also the largest.

Nonlinear corrections to the radii of large-amplitude pulsating stars have not been systematically studied in the literature so far. It is therefore essential to investigate how this effect depends on pulsation mode, pulsation amplitude, fundamental stellar parameters, and the treatment of pulsation–convection coupling. Moreover, the recent result of \cite{Farag-2026}, although based on a single Cepheid model, suggests that the radius variation may be sensitive to the numerical solver used to integrate the nonlinear pulsation equations. \cite{Farag-2026} proposed a self-consistent calculation of stellar pulsation based on envelope models fully consistent with the underlying evolutionary structure. Such computations are better suited to address the problem of nonlinear radius corrections; however, they are computationally demanding and beyond the scope of the present analysis. A follow-up study is planned; see also Sec.~\ref{sec:discussion}.

We also note that once radii are determined with a precision comparable to that achieved for our F-mode sample, additional effects begin to play a role. Although Cepheids are relatively slow rotators, their radii may still be affected by rotation -- an effect that has not been taken into account so far as well. Also the precise definition of what stellar radius is, starts to matter. We discuss more in Sect.~\ref{sec:discussion}.

At present, we conclude that the pulsation period should be the primary constraint used when matching models to observations of classical Cepheids. A consistent use of the stellar radius requires a dedicated study to quantify the role of nonlinear radius corrections and other effects not considered so far (see also Sect.~\ref{sec:discussion}). In the following section, solutions for Cepheids in the final model grid are selected based on \chis{}, including \Teff{}, $L$, and pulsation period.

\subsection{Solutions for final model grid}\label{ssec:gridtwo}

This section presents solutions for the final model grids introduced in Sects.~\ref{ssec:grids} and \ref{ssec:gridone}. Mass loss is included, and the stellar masses are allowed to vary within the observational errors from Tab.~\ref{tab:dlebs}. In the \chis{}, $\log L$, $T_{\rm eff}$, $P$ are included for the Cepheid, and $\log L$, $T_{\rm eff}$, $R$, for the non-pulsating stars. Additional observational constrains are also taken into account.

While the composition of the $\chi^2$ differs between Cepheid and non-pulsating components, with the pulsation period entering only for the former and the radius only for the latter, this asymmetry is physically motivated. The issue of nonlinear radius corrections is specific to pulsating stars and therefore does not apply to non-pulsating components. In principle, the solution for non-pulsating stars could be based solely on $L$ and $T_{\rm eff}$; however, excluding the radius would discard the most precisely determined observable for these stars and thus weaken the constraints provided by the binary system.

\begin{figure}
\centering
\includegraphics[width=\linewidth]{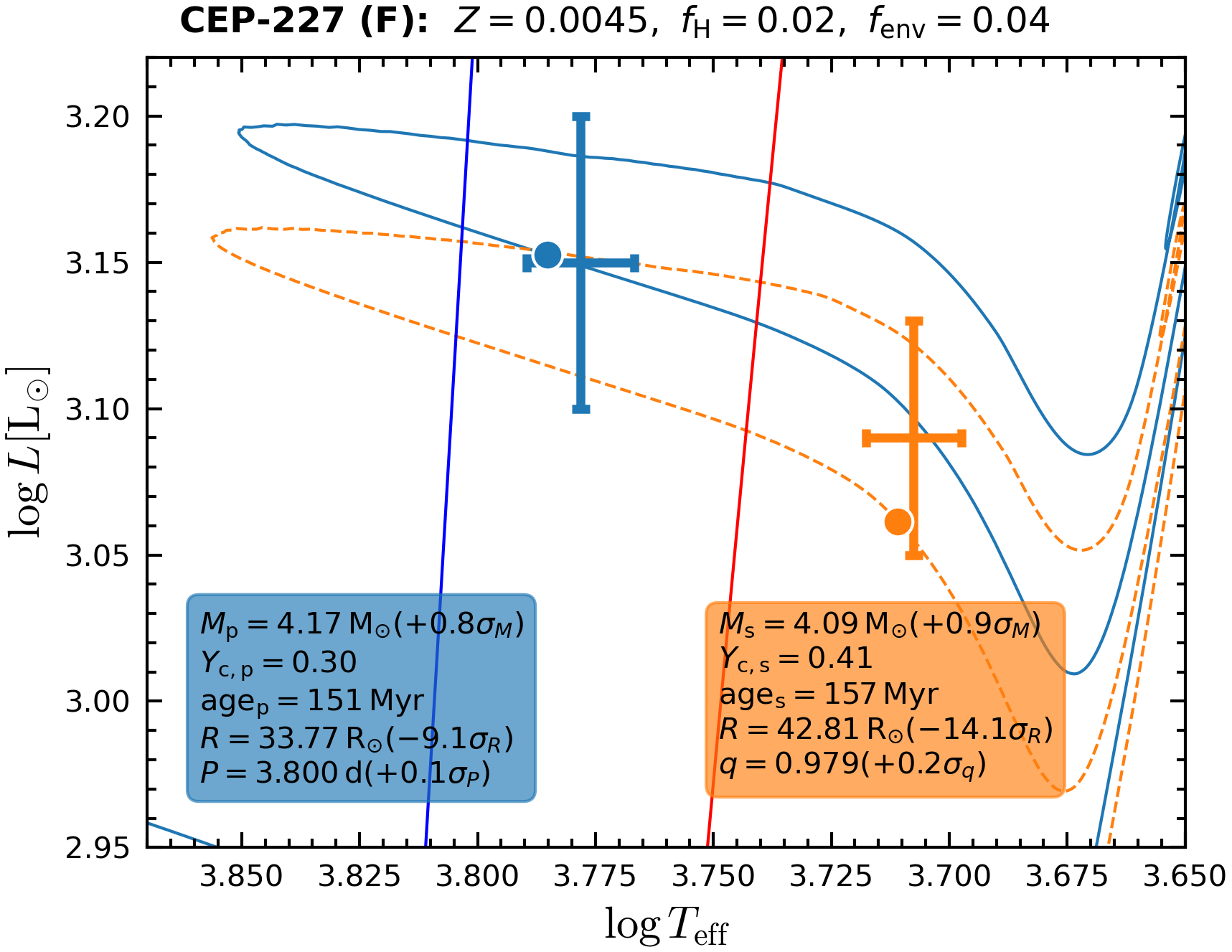}
\caption{Best-matching solution for CEP-0227 in the final grid.}
\label{fig:227_2}
\end{figure}

For CEP-0227 (Fig.~\ref{fig:227_2}; parameters of the solutions presented in the figure and in more detail in Tab.~\ref{tab:sol-0227}), the less massive companion is forced onto the lower branch of the loop, so it is no more advanced than the Cepheid. The metallicity, $Z$=0.0045, is slightly higher than in the initial solution (Fig.~\ref{fig:grid1B}), while \fcor=0.02 remains the same. The masses are higher than the central values, but still within the observational errors, and the same applies to the mass ratio. The age difference between the two stars, 6\,Myr, is significantly smaller than in the previous solutions. The Cepheid's radius is  9$\sigma_R$ away from the observed value. Applying the nonlinear radius correction (Sect.~\ref{ssec:pvsr}) reduces the discrepancy to 6$\sigma_R$, but it remains significant. Otherwise, the solution is satisfactory; mass inversion is not required.

\begin{figure}
\centering
\includegraphics[width=\linewidth]{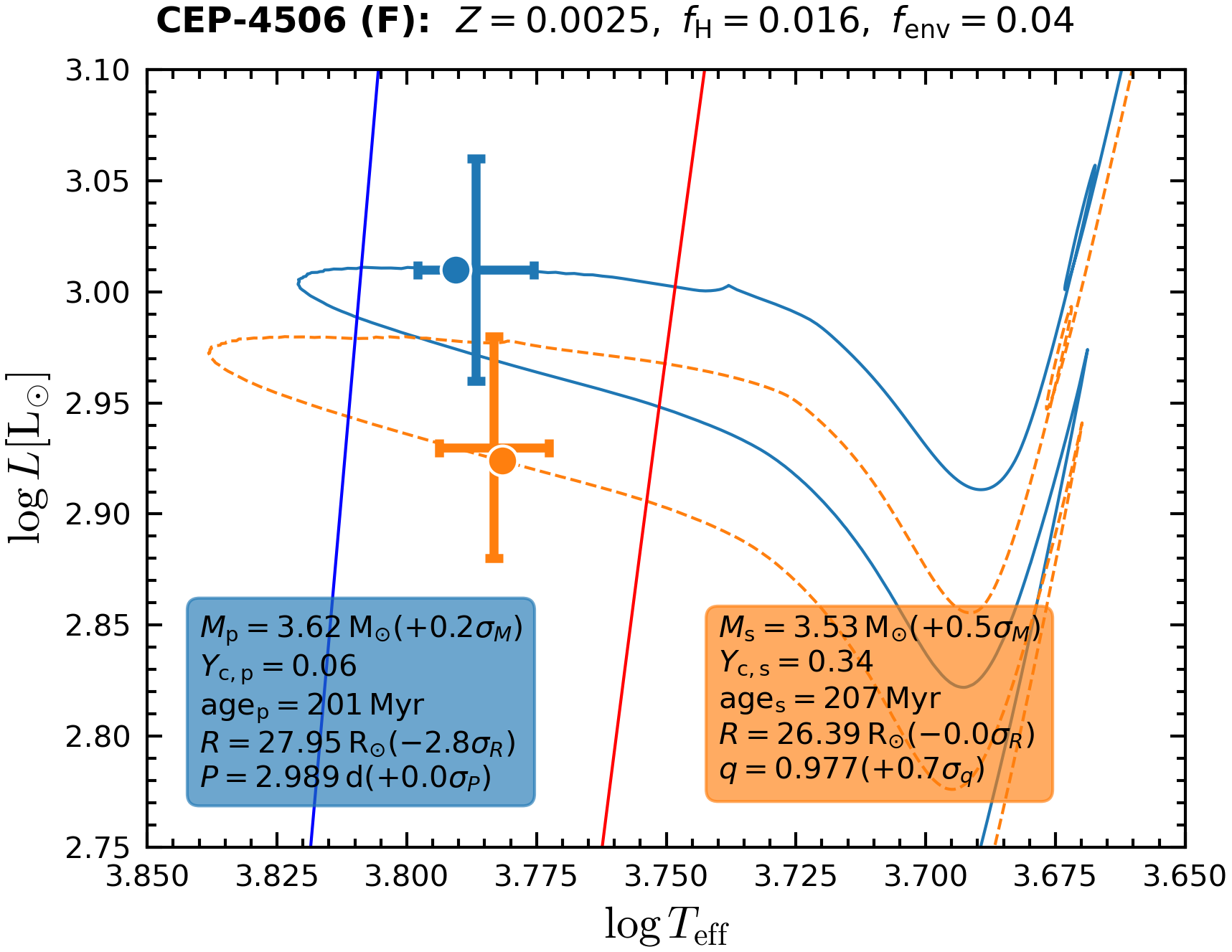}
\caption{Best-matching solution for CEP-4506 in the final grid.}
\label{fig:4506_2}
\end{figure}

For CEP-4506 (Fig.~\ref{fig:4506_2}, Tab.~\ref{tab:sol-4506}), the solution we obtain is qualitatively the same as that presented in Sect.~\ref{ssec:gridone} (Fig.~\ref{fig:grid1B}), with the Cepheid found on the third crossing and the non-pulsating star on the second crossing. The masses are $0.5\sigma_M$ higher, and the mass ratio is in agreement with the measured value. An equally good solution may be found without increasing the masses. The metallicity is $Z$=0.0025, slightly lower than in the initial solution, but the overshooting efficiency is also lower, \fcor=0.016. The age difference, 6\,Myr, is reasonable. As expected, the Cepheid's radius is 3$\sigma_R$ away from the observed value.

In the Appendix, Fig.~\ref{fig:4506_2b} (Tab.~\ref{tab:sol-4506_alt}) presents a qualitatively similar solution, but with a slightly higher metallicity, $Z$=0.0035 (which decreases the loop's luminosity), and a slightly larger \fcor=0.02 (which increases the loop's luminosity). This illustrates the degenerate effects of varying $Z$ and \fcor, and highlights the importance of spectroscopic metallicity constraints.

\begin{figure}
\centering
\includegraphics[width=\linewidth]{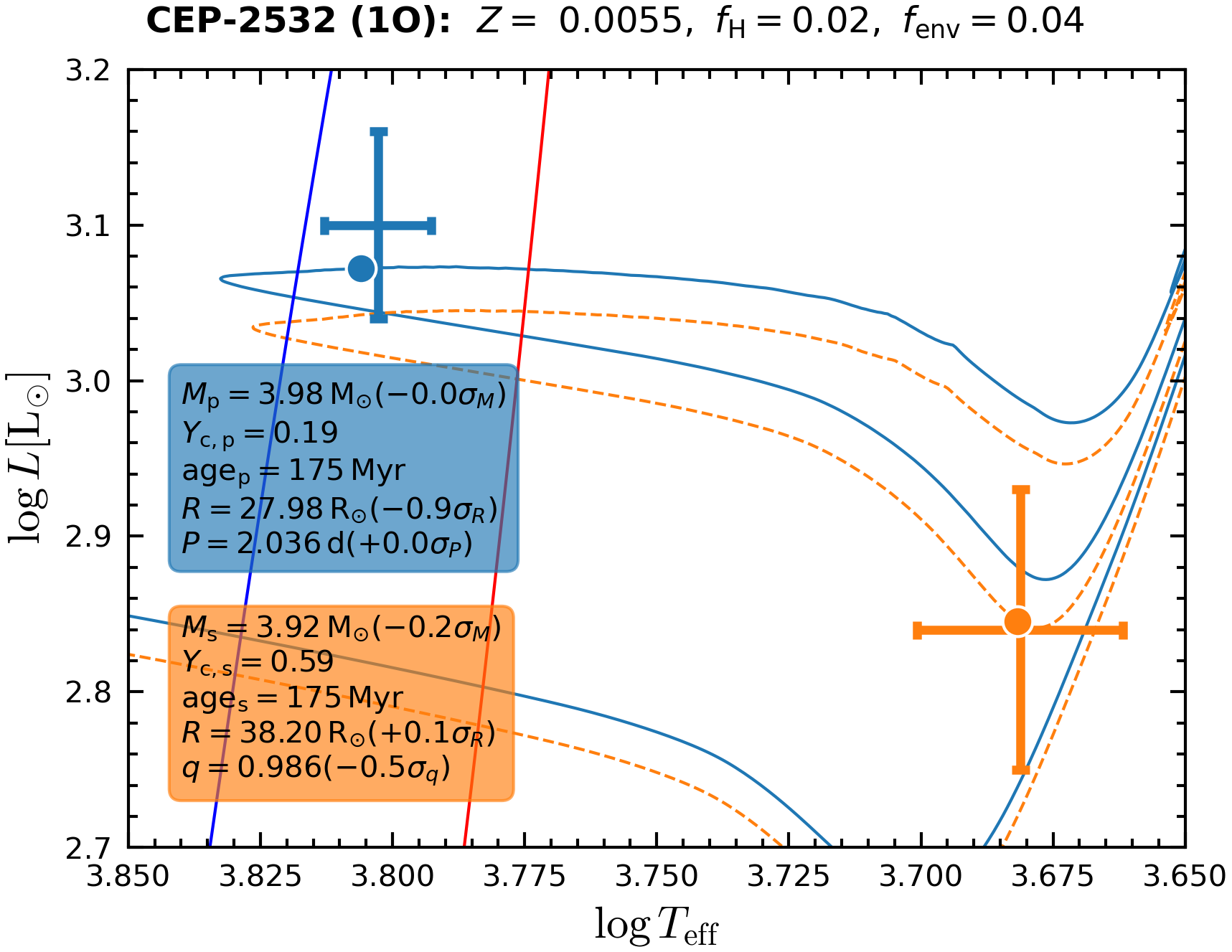}
\caption{Best-matching solution for CEP-2532 in the final grid.}
\label{fig:2532_2}
\end{figure}

For CEP-2532 (Fig.~\ref{fig:2532_2}, Tab.~\ref{tab:sol-2532}), the final solution is very similar to the initial one (with $Z$=0.006, Fig.~\ref{fig:grid1}). The Cepheid is found on the third crossing, and the companion on the lower branch of the blue loop, at the early stage of the CHeB. The system is modeled with $Z$=0.0055 and \fcor=0.02, but other $Z$–\fcor{} combinations also yield satisfactory solutions; again, only spectroscopic determination of metallicity may lift this degeneracy. The ages match, the mass ratio is within the observational error, and both radii are also consistent with observations. An alternative solution, with a larger age difference, exists for the Cepheid on the second crossing and the companion on the RGB (Fig.~\ref{fig:2532_2b}, Tab.~\ref{tab:sol-2532_alt}).

\begin{figure}
\centering
\includegraphics[width=\linewidth]{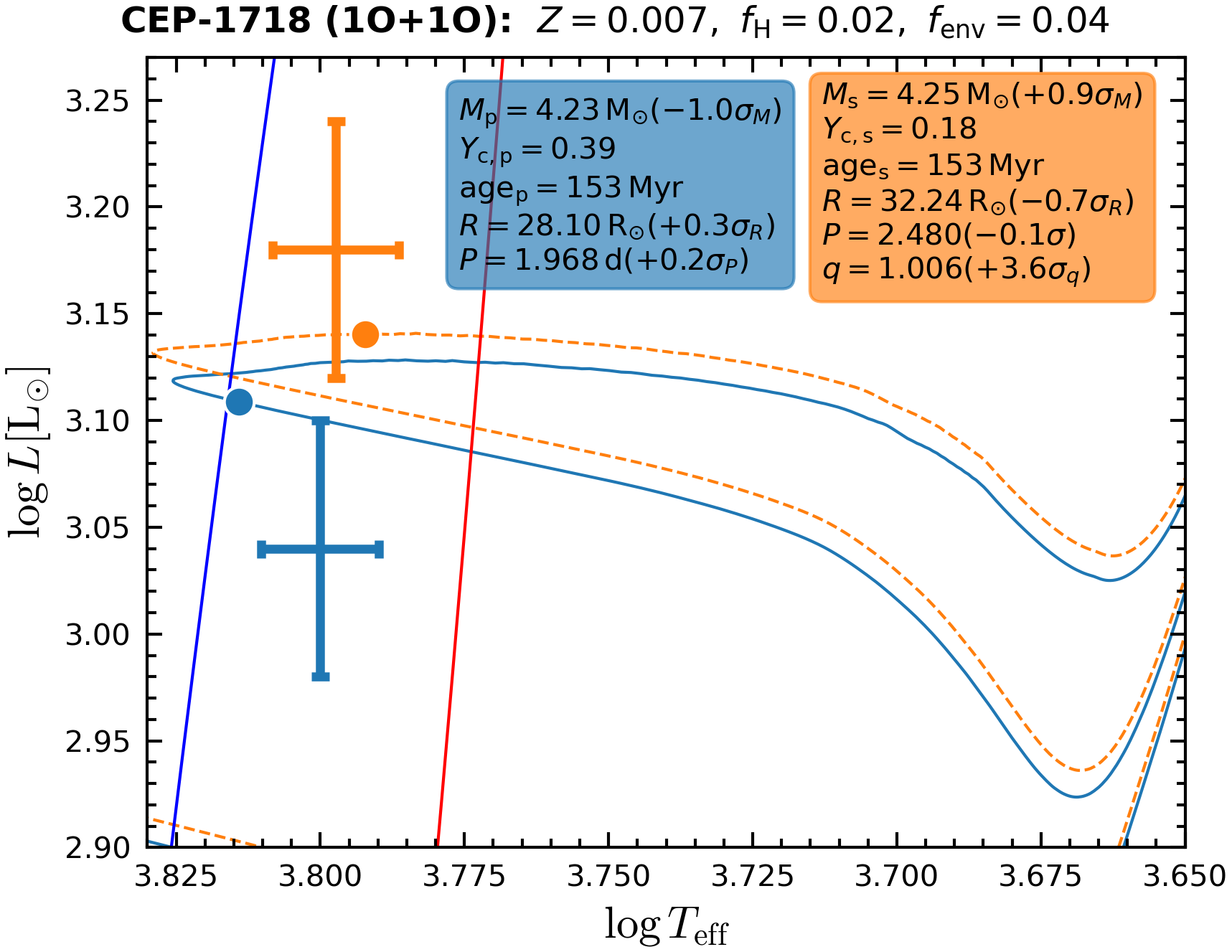}
\caption{Best-matching solution for CEP-1718 in the final grid with mass inversion and the PCR condition lifted for the secondary.}
\label{fig:1718_2}
\end{figure}

For the final modeling of CEP-1718 (Fig.~\ref{fig:1718_2}, Tab.~\ref{tab:sol-1718}), we find a more reasonable solution assuming a mass inversion scenario. In this case, however, the mass ratio, 
$q$=1.006, is 3.5$\sigma_q$ away from the measured value. The secondary, now more massive star, is then found on the third crossing (despite the PCR measurement), and the primary on the second crossing, close to the turning point of the loop. The ages of the two Cepheids are the same, and the radii are in agreement with the observed values. In this scenario, 
$Z$=0.007 and \fcor=0.02. The PCR constraint is not satisfied, as the secondary evolves redward. In the Appendix (Fig.~\ref{fig:1718_2b}, Tab.~\ref{tab:sol-1718_alt}) we show a solution based on the same evolutionary tracks, in which the secondary is forced on the second crossing, in agreement with PCR.

CEP-1718 is clearly the most challenging system in our sample to model, owing to the combination of observational constraints: both components have similar effective temperatures, while the more massive component appears to be less luminous. For this reason, we explored a mass-inversion solution despite its mass ratio being $3.5\sigma_q$ away from the observed value. At the same time, the system is observationally challenging, as only a single shallow eclipse is present in the light curve, which may limit the precision with which some parameters can be determined. It is therefore difficult to assess whether additional observations could lead to a revision of the mass ratio. We also note that formal observational uncertainties may occasionally underestimate the true errors. As an example, for the extensively observed system CEP-227, the mass ratio evolved from $q=1.00\pm0.01$ \citep{Pietrzynski-2010} to $q=0.993\pm0.002$ \citep[][RAVESPAN solution]{Pilecki-2013}, and finally to $q=0.979\pm0.003$ \citep{Pilecki-2018}. The latter two determinations differ by nearly $4\sigma$ when the quoted uncertainties are adopted. While this example does not demonstrate that the mass ratio of CEP-1718 is incorrect, it illustrates that formally significant discrepancies can arise as observational analyses improve. Nevertheless, the mass-inversion solution should be regarded with caution until stronger observational constraints become available.

For all systems except CEP-1718, the best solutions require lower $Z$ than the assumed LMC average. Evolutionary models exhibit many degeneracies -- for example, a lower $Z$ or a higher \fcor{} both increase luminosity. A similar effect is obtained by including other types of extra mixing, such as rotational mixing, or by increasing the stellar mass. The systematically lower $Z$ in the best-fit solutions does not necessarily imply that these stars actually have lower $Z$ than their host galaxy; rather, it may reflect the mass discrepancy problem. Until more precise constraints on $Z$ are available, it remains unclear whether the systems' metallicity is genuinely lower or whether enhanced mixing is required.

The presented systems were also analyzed by \cite{Deka-2025}, who included $L$, \Teff{} and $R$ in \chis{} and, for most stars, considered only two metallicities, [Fe/H]=$-0.5$ and $-0.6$. In their models, MS core overshooting follows from the turbulent convection prescription of \cite{Ahlborn-2022} and is not parametrized. Although a direct comparison is not possible, the solutions and reported issues are similar. For CEP-0227, their [Fe/H]=$-0.5$ solution closely matches that shown in Fig.~\ref{fig:grid1}, with both components on the third crossing. For correct evolutionary ordering of the two components, they invoke an inverted mass ratio. For CEP-4506, the adopted metallicities are too high, resulting in blue loops that are too dim relative to the observed HRD positions, and both components are placed on the third crossing. For CEP-2532, they also present alternative solutions, with the Cepheid on the blue loop and the secondary either at the beginning of the CHeB or on the RGB. For CEP-1718, they likewise note the intrinsic difficulty of modeling the system and present three scenarios with both components at different evolutionary stages, none reproducing the correct ordering, as mass inversion is not considered.

\section{Discussion and conclusions}\label{sec:discussion}

We have investigated whether the pulsation period -- the most precisely determined observable for classical Cepheids in eclipsing binary systems -- can be used to constrain evolutionary models. To this end, we modeled five classical Cepheids in DLEBs with accurately determined physical parameters from \citetalias{Pilecki-2018}. We used $\chi^2$ minimization to select the best-fitting evolutionary models computed with \MESA.

We find that satisfactory solutions can be obtained when matching effective temperature, luminosity, and radius, or when matching effective temperature, luminosity, and pulsation period, but not radius and pulsation period simultaneously (Sect~\ref{ssec:pvsr}, Fig.~\ref{fig:PRtensionFO}). For F-mode Cepheids, when the pulsation period is matched, the predicted radii are systematically too small, by $3$ to $12\sigma_R$, where $\sigma_R$ denotes the observational uncertainty. This tension is revealed thanks to the combination of very precise ($<$1\%) radius determinations and comparably precise ($\sim$1\%) pulsation periods derived from models. In contrast, the uncertainties in effective temperature and luminosity are significantly larger, allowing both radius- and period-based solutions to fall within the same error box in the HRD. While this discrepancy is clear for F-mode Cepheids, it is less evident for 1O Cepheids, primarily because their radii are determined with an order of magnitude lower precision. 

When matching models and observations, it is essential to compare corresponding quantities. The pulsation period is straightforward. Observationally it is measured with extremely high precision and is free of systematic effects. Its determination from models is also straightforward, and the only subtlety lies in the fact that the nonlinear period should be compared with observations. We addressed this by applying nonlinear period corrections determined for the modeled systems.

The situation is different for the radius. Observationally, its derivation in eclipsing binary systems is more complex \citep[see][]{Pilecki-2013}, as it is based on modeling the light and radial velocity curves and depends on the adopted orbital solution. Consequently, several factors contribute to both the statistical and systematic error budget. On the theoretical side, the static radius associated with \MESA{} evolutionary tracks corresponds to an optical depth of $\tau$=2/3, and the three quantities, $L$, \Teff{}, and $R$, are related by $L=4\pi R^2\sigma\Teff^4$. We note there is no unique definition of stellar radius, and while different definitions should be consistent for compact stars, it is not necessarily the case for stars with extended envelopes \citep{Baschek-1991}. 

The radius at full-amplitude nonlinear pulsation, which may be computed with \RSP, is larger than the static one. The nonlinear correction, while a small fraction of the radius itself, may exceed the precision of the observational determination. When taken into account, it significantly decreases the discrepancy for most of the studied Cepheids. The exception is CEP-0227, for which the discrepancy remains significant and additional factors must contribute.

Apart from unaccounted systematic errors, either in the observations or in the modeling, the effects of rotation should also be investigated in more detail. Even though Cepheids are slow rotators \citep[e.g.,][]{Nardetto-2006, Anderson-PhD}, rotation may contribute to explaining the remaining discrepancies.

The calculations of pulsation periods with \gyre{}, presented in Sect.~\ref{ssec:puls}, also indicate that there may be a systematic effect in the computed periods (which would further reduce the discrepancy), although it appears to be small. In Appendix~\ref{secapp:RobustTension} we show that when assuming the largest shift in linear periods that we computed for our sample, namely 1.4\%, as a systematic error on the linear pulsation periods, the radius tension remains significant and is only slightly reduced.

We note that the radius discrepancy problem is well known for low-mass (0.1--1.25\MS) stars \citep[e.g.,][]{Torres-2010}, where radii are under-predicted by stellar models by about 3\% \citep{Spada-2013}. For this mass range, radius inflation is connected with increased magnetic activity \citep[e.g.,][]{Chabrier-2007}.

The issue of nonlinear radius correction lacks systematic study. Such a study must address its dependence on pulsation mode, the basic physical parameters of the model, and the parameters describing pulsation–convection coupling. Moreover, recent improvements in time-dependent convection (TDC) in \MESA{} \citep{Farag-2026} suggest that the results may be sensitive to the numerical solver used in nonlinear calculations. The new TDC scheme also allows nonlinear pulsation models to be computed directly on top of evolutionary calculations with exactly the same model structure and physical assumptions, which would allow us to study the problem in a fully self-consistent way. Conducting such a study is very time-consuming and beyond the scope of the present work, but it is planned. We also intend to assess the impact of rotation on Cepheid radii as well as to exploit other pulsation codes, in particular \gyre{}, to investigate possible systematics in the computation of pulsation periods.

We also note that nonlinear radius corrections have a much broader impact than modeling eclipsing binaries alone. Period--radius (PR) relations derived based on static evolutionary models \citep[e.g.,][]{Anderson-2016, Smolec-2026} should be corrected for nonlinear effects before comparison with observations. Relations based on nonlinear pulsation models \citep[e.g.,][]{Bono-1998, DeSomma-2022} are free of this issue. Nevertheless, the expected nonlinear correction to the radius is very small, smaller than the precision of most radius determinations available in the literature and significantly smaller than the scatter of published PR relations. The latter often do not include an explicit color term and are based on specific assumptions regarding metallicity, overshooting, or the underlying mass--luminosity relation. For these reasons, published PR relations are unlikely to reveal the tension discussed in this paper or provide strong constraints on its interpretation. We illustrate this point in the Appendix~\ref{secapp:pr}. While PR relations are valuable tools for studies of Cepheid populations, for analyses of moderate and large samples of stars, and for testing stellar evolution and pulsation theory \citep[e.g.,][]{Anderson-2016, DeSomma-2022, Bailleul-2026, Smolec-2026}, dedicated modeling of individual stars is required when radii are determined with subpercent precision, as is the case for the systems analyzed in this work.

With precise determination of theoretical pulsation periods, it is tempting to use them directly in the process of determining the physical parameters of classical Cepheids in eclipsing binary systems, as this should allow for tightening of the uncertainties in their position on the HRD, both in effective temperature and absolute luminosity. This expectation stems from the fact that in the HRD, the region in which the pulsation period is matched constitutes only a fraction of the current error box (Fig.~\ref{fig:PRtensionFO}). A similar argument applies to the stellar radius, once all systematic factors on the theoretical side are properly settled and quantified.

We plan to conduct a systematic study of nonlinear radius corrections in future work. For the moment, we suggest that the pulsation period should be used to constrain evolutionary models of Cepheids in eclipsing binary systems.

We demonstrated that satisfactory solutions may be obtained for most of the studied systems when Cepheid models are constrained by the pulsation period (Sect.~\ref{ssec:gridtwo}). We note that intrinsic degeneracies are present in the modeling. The most important is that between metallicity and the efficiency of MS core overshooting. Lowering metallicity and increasing core overshooting both increase the luminosity of the blue loop. Consequently, the determination of metallicity -- which is currently lacking for all considered systems -- is essential for constraining evolutionary models and for studying the mass discrepancy problem in the context of eclipsing binaries. We also require significantly more precise determinations of other observables. In particular, more accurate radii for 1O Cepheids would allow us to study nonlinear effects on radii for both pulsation modes.

We also note that some systems appear intrinsically difficult to reconcile with observations. The best example is CEP-1718, in which the less massive component is significantly brighter while having nearly the same effective temperature as the more massive, dimmer component. A physically realistic solution for this system that does not invoke mass inversion cannot be found.

\begin{acknowledgements}
This research is supported by the National Science Center, Poland, Sonata BIS project 2018/30/E/ST9/00598. We thank the referee, Giulia De Somma, for helpful and constructive comments that improved the quality of this paper.
\end{acknowledgements}

\bibliography{main_aa} 
\bibliographystyle{aa}

\begin{appendix}
\nolinenumbers

\section{Physical properties of Cepheids and their companions in DLEBs}\label{secapp:tabelka}

In Tab.~\ref{tab:dlebs} we collect the characteristics of the modeled DLEBs from \citetalias{Pilecki-2018}.

\begin{table*}[!t]
\caption{Physical properties of Cepheids and their companions in DLEBs in the LMC from \citetalias{Pilecki-2018}.}
\label{tab:dlebs}
\centering
\begin{tabular}{llllllll}
\hline\hline
name & $P$ (d) & mode  & $M$ (\MS) & $q$ & $R$ (\RS) & $T_{\rm eff}$ (K) & $\log L$ (\LS) \\ 
\hline
CEP-0227  & 3.797086 &  F & 4.15 $\pm$ 0.03 & 0.979 $\pm$ 0.003  & 34.87 $\pm$ 0.12  & 6000 $\pm$ 160 & 3.15 $\pm$ 0.05 \\ 
          &          &    & 4.06 $\pm$ 0.03 &                    & 44.79 $\pm$ 0.14  & 5100 $\pm$ 120 & 3.09 $\pm$ 0.04 \\ \hline
CEP-4506  & 2.987846 &  F & 3.61 $\pm$ 0.03 & 0.975 $\pm$ 0.003  & 28.5  $\pm$ 0.2   & 6120 $\pm$ 160 & 3.01 $\pm$ 0.05 \\
          &          &    & 3.52 $\pm$ 0.03 &                    & 26.4  $\pm$ 0.2   & 6070 $\pm$ 150 & 2.93 $\pm$ 0.05 \\ 
\hline
CEP-1812  & 1.312903 &  F  & 3.76 $\pm$ 0.03 & 0.696 $\pm$ 0.003 & 17.85 $\pm$ 0.13  & 6120 $\pm$ 150 & 2.61 $\pm$ 0.04 \\
          &          &     & 2.62 $\pm$ 0.02 &                   & 11.83 $\pm$ 0.08  & 5170 $\pm$ 120 & 1.95 $\pm$ 0.04 \\
\hline 
CEP-2532  & 2.035349 &  1O & 3.98 $\pm$ 0.10 & 0.992 $\pm$ 0.012 & 29.2  $\pm$ 1.4   & 6350 $\pm$ 150 & 3.10 $\pm$ 0.06 \\
          &          &     & 3.94 $\pm$ 0.09 &                   & 38.1  $\pm$ 1.8   & 4800 $\pm$ 220 & 2.84 $\pm$ 0.09 \\ \hline
CEP-1718A & 1.963663 &  1O & 4.27 $\pm$ 0.04 &                   & 27.8  $\pm$ 1.2   & 6310 $\pm$ 150 & 3.04 $\pm$ 0.06 \\
CEP-1718B & 2.480917 &  1O & 4.22 $\pm$ 0.04 & 0.988 $\pm$ 0.005 & 33.1  $\pm$ 1.3   & 6270 $\pm$ 160 & 3.18 $\pm$ 0.06 \\
\hline
\end{tabular}
\tablefoot{ Columns contain the Cepheid's OGLE identifier, pulsation period, pulsation mode, mass, mass ratio, $q$, radius, effective temperature, and absolute luminosity.}
\end{table*}

\section{Treatment of mass loss}\label{secapp:massloss}

In our calculations in Sect.~\ref{ssec:gridtwo}, we assume mass loss on the RGB and during the CHeB following the Reimers formula with $\eta$=0.4. We note that the efficiency of mass loss does not noticeably affect the evolutionary track in the HRD \citep{Smolec-2026}. It primarily determines the amount of mass lost at a given evolutionary stage, which also depends on mass, metallicity, and overshooting parameters.

Our initial grid of models (Sect.~\ref{ssec:gridone}) is computed with the exact masses from Tab.~\ref{tab:dlebs}, with mass loss neglected. We then examine the solutions and adjust the initial mass in the final, more detailed grid of models -- including mass loss -- to recover the correct mass at the expected evolutionary stage (Sect.~\ref{ssec:gridtwo}). To calibrate the necessary corrections, we first investigate how the fraction of mass lost by the star depends on its mass, metallicity, overshooting parameters, and evolutionary stage.

\begin{figure}[h]
    \centering
    \includegraphics[width=\linewidth]{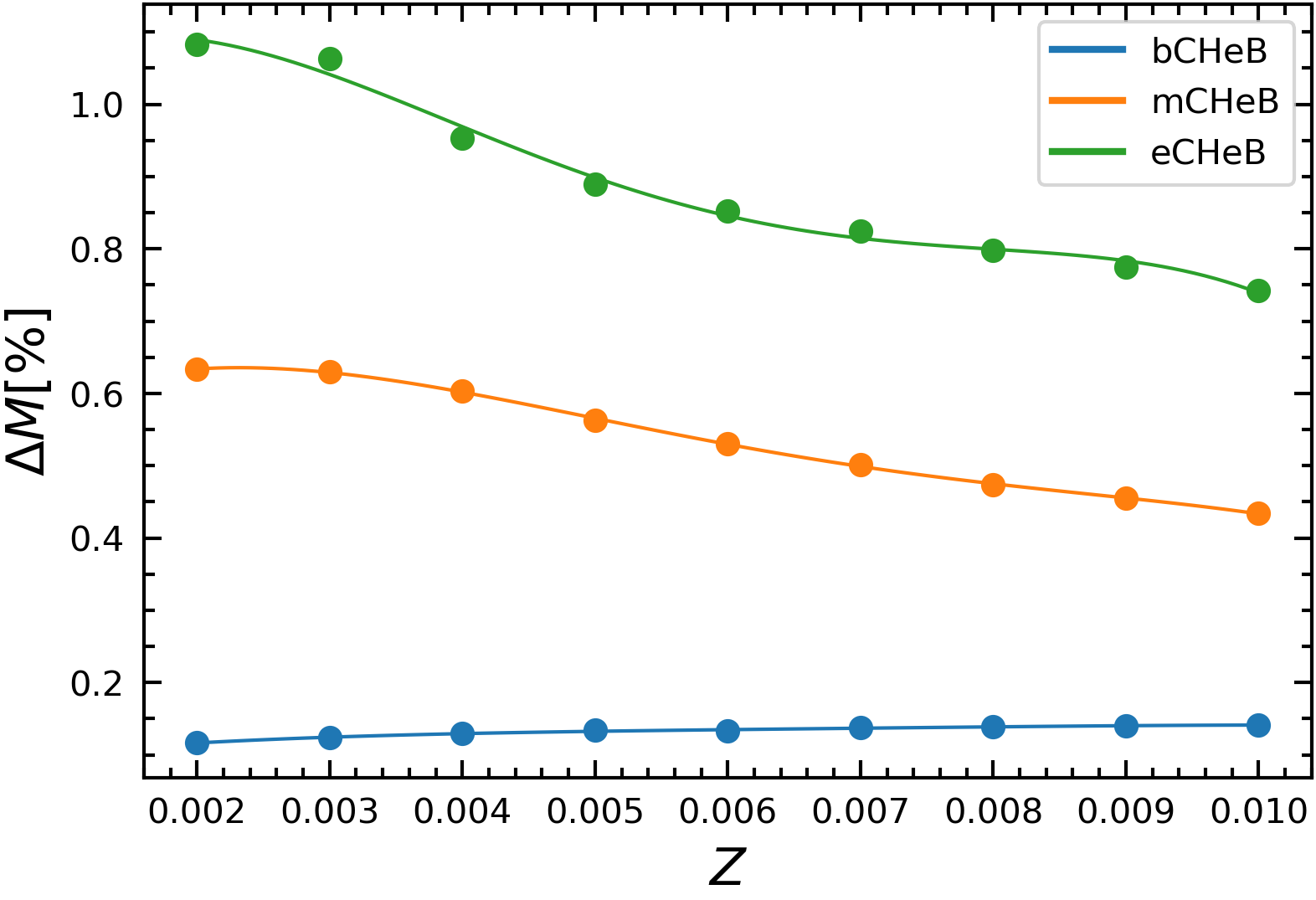}
    \caption{Relative mass lost, $\Delta M=(M-M_0)/M_0$, as a function of metallicity (x-axis) and evolutionary stage (color). All tracks have the same mass and overshooting ($M$=4\MS, \fcor=0.02, \fenv=0.04). A fitted polynomial of 4th degree is over-plotted.}
    \label{fig:dMZ}
\end{figure}

First, we calculate models with 4\MS{} (all modeled systems have masses close to 4\MS), moderate overshooting from the core (\fcor=0.02) and envelope (\fenv=0.04), and a range of metallicities $Z$=0.002-–0.010 with a step of $dZ$=0.001. The percentage of mass lost by the beginning, middle, and end of the core-helium-burning phase \citep[bCHeB, mCHeB, eCHeB; as defined in][]{Ziolkowska-2024} is presented in Fig.~\ref{fig:dMZ}. The dependence on metallicity is weak, but it increases as helium burning in the core advances. By the eCHeB, a low-metallicity ($Z$=0.002) star loses 1.1\% of its initial mass, while a higher-metallicity ($Z$=0.010) star loses 0.75\%. What is important is that, at a given evolutionary stage and for metallicities close to $Z$=0.006, the dependence of mass loss on metallicity is weak.

\begin{figure}[h]
    \centering
    \includegraphics[width=\linewidth]{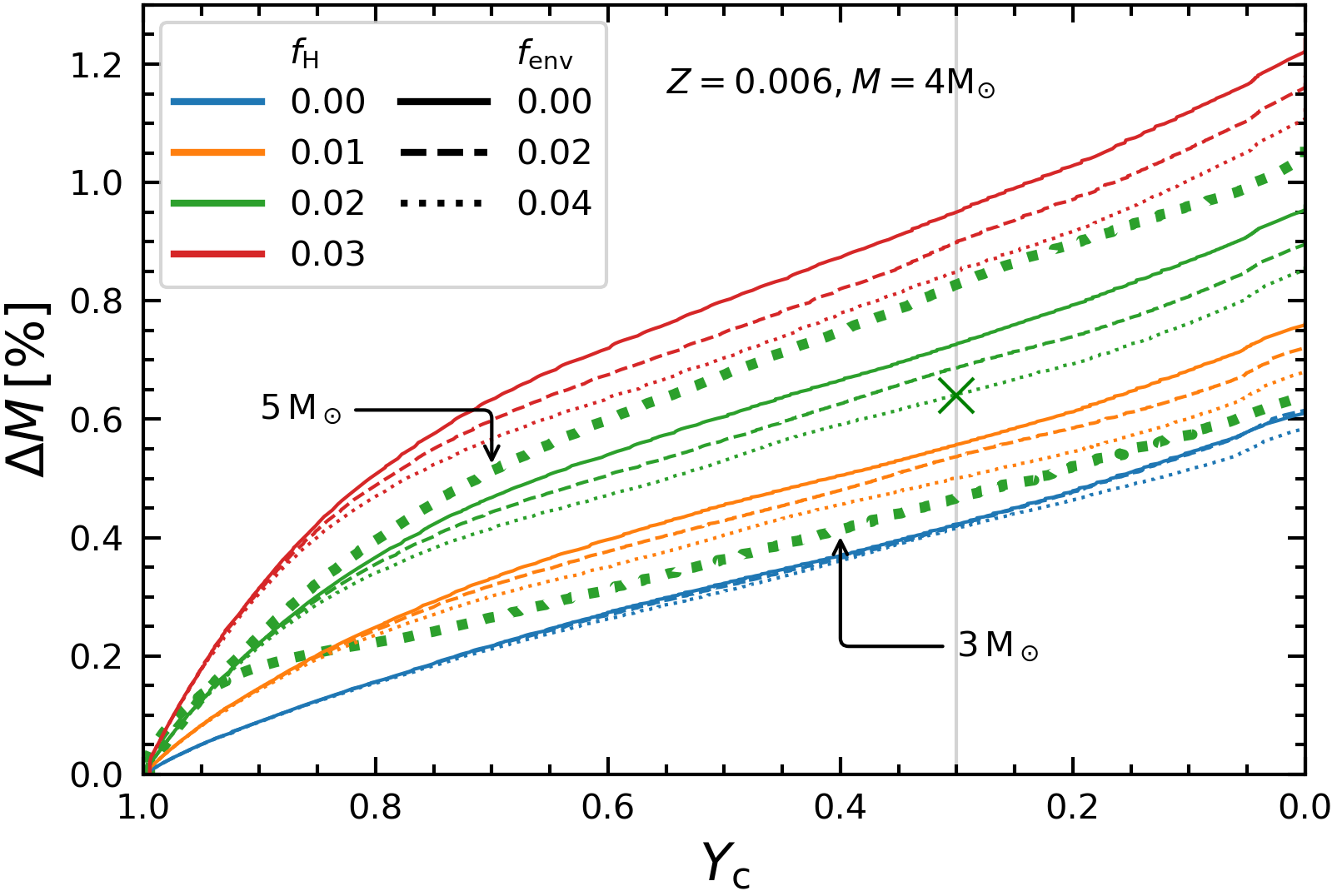}
    \caption{Relative mass lost, $\Delta M=(M-M_0)/M_0$, as a function of central helium content. Models plotted adopt different overshooting efficiency from the MS core (color), envelope (line styles). Only post-MS evolution is plotted. All models have 4\MS{} and $Z$=0.006, except for the two tracks marked by the thick dotted line which have different masses, 3 and 5\MS, with \fcor=0.02 and \fenv=0.04.}
    \label{fig:ml}
\end{figure}

In Fig.~\ref{fig:ml} we present the relative mass lost as a function of central helium content (a measure of evolutionary time during the CHeB) for 4\MS{} models with $Z$=0.006, computed for different core (colors) and envelope (line styles) overshooting efficiencies. The stronger the core overshooting, the more mass is lost -- up to a 0.6\% difference between no overshooting and strong overshooting by the eCHeB. The mass loss does not depend strongly on envelope overshooting. We also include two tracks with 3\MS{} and 5\MS{} in Fig.~\ref{fig:ml}, computed for the same $Z$ and \fcor=0.02, \fenv=0.04 (green thick dotted lines), to show that changing the initial mass has a similar, though slightly weaker, effect than changing the efficiency of overshooting.

Our goal is not to precisely reproduce the observed masses, as their observational errors (see Tab.~\ref{tab:dlebs}) are of the same order as the expected mass loss. Hence, we adopt a simplified approach and apply a constant correction. Taking advantage of the fact that, at a given evolutionary stage, the mass lost by the star depends only weakly on mass, metallicity, and overshooting parameters, and that the masses of all modeled LMC systems are close to 4\MS{}, the corrections are based on a single track (4\MS, $Z$=0.006, \fcor=0.02, \fenv=0.04; thin, green dotted line in Fig.~\ref{fig:ml}. An inspection of Figs.~\ref{fig:grid1} and \ref{fig:grid1B}, together with the values of central helium content for the best-matching solutions, leads to a "universal" correction for all blue loop systems corresponding to  $Y_{\rm c}$=0.3 (marked with cross in Fig.~\ref{fig:ml}). While this is a strong simplification, it allows us to recover the observed masses reasonably well, and the impact of mass loss on evolutionary tracks, although very weak, is explicitly taken into account in Sect.~\ref{ssec:gridtwo}.

\section{Robustness of the period--radius tension}\label{secapp:RobustTension}

\begin{figure*}
\centering
\includegraphics[width=\linewidth]{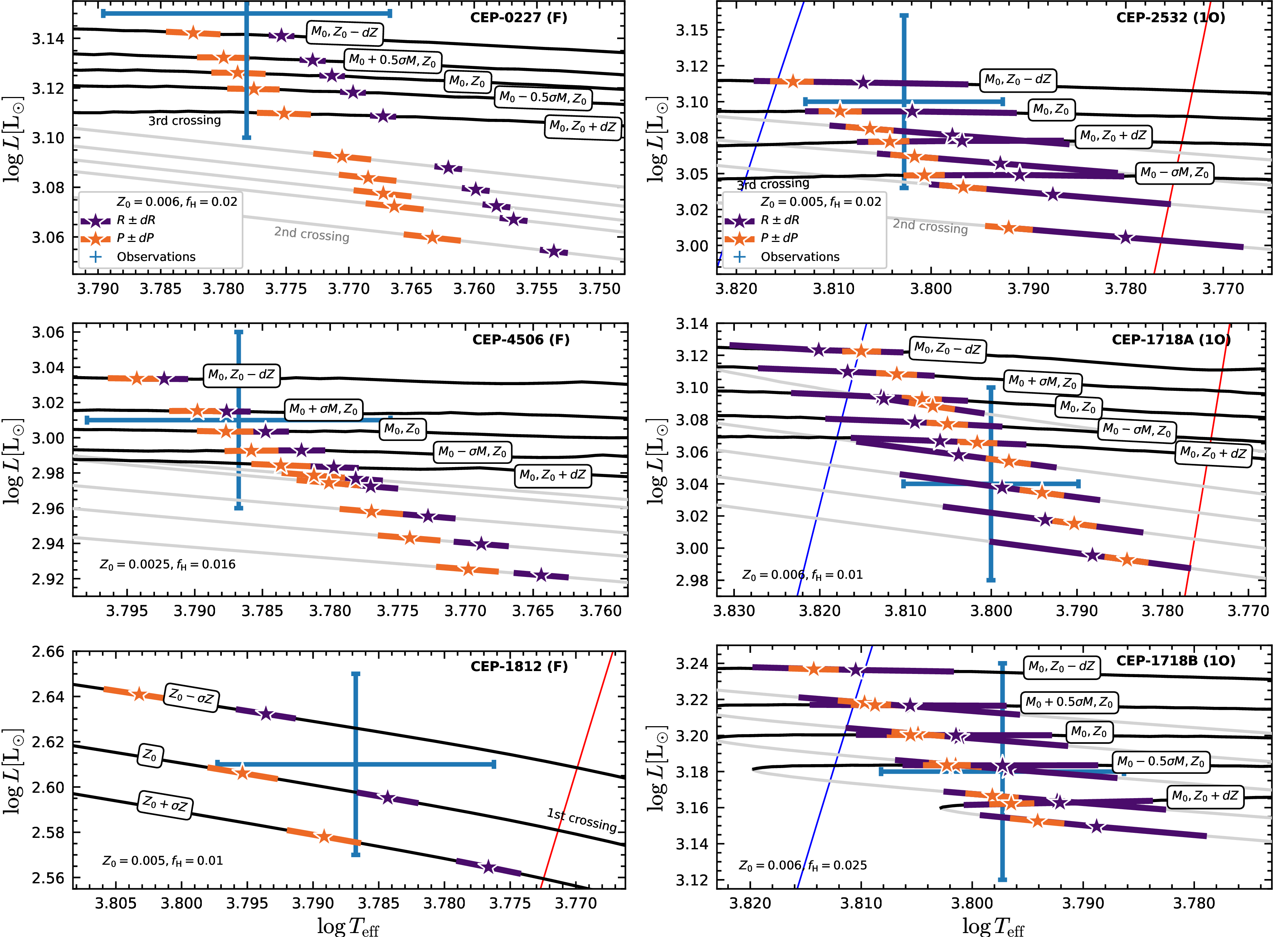}
\caption{A variant of Fig.~\ref{fig:PRtensionFO} in which we assume a larger uncertainty of the pulsation periods of 1.5\% and their systematic shift by 1.4\% toward shorter values.}
\label{fig:PRtension_mod}
\end{figure*}

The significance and magnitude of the tension between solutions based on matching the pulsation period and those based on matching the radius, illustrated in Fig.~\ref{fig:PRtensionFO}, depend on the adopted assumptions regarding the uncertainty in the pulsation periods and any possible systematic errors in their determination. To demonstrate that the discrepancy is significant, we prepared a modified version of Fig.~\ref{fig:PRtensionFO}, shown in Fig.~\ref{fig:PRtension_mod}. In constructing this figure, we increased the uncertainty in the pulsation periods to 1.5\% (nearly three times the value estimated in Sect.~\ref{ssec:puls}) and assumed a systematic shift in the computed linear periods. At a given position in the HR diagram, we assume that the pulsation period is 1.4\% shorter than that predicted by the \RSP{} calculations. This corresponds to the largest difference between the periods computed with \RSP{} and \gyre{} (Sect.~\ref{ssec:puls}).

In practice, only the systematic shift in the pulsation period affects the discrepancy between the observed radius and the radius determined at the location where the pulsation period of the Cepheid is reproduced. Under these assumptions, the tension is indeed reduced, but only to a limited extent. For the individual Cepheids, the changes are as follows. For the F-mode Cepheids, the tension decreases from $-12.0\sigma_R$ to $-9.9\sigma_R$ (CEP-0227), from $-3.2\sigma_R$ to $-2.1\sigma_R$ (CEP-4506), and from $-6.2\sigma_R$ to $-5.2\sigma_R$ (CEP-1812). For the 1O Cepheids, the corresponding changes are from $-0.9\sigma_R$ to $-0.8\sigma_R$ (CEP-2532), from $0.3\sigma_R$ to $0.5\sigma_R$ (CEP-1718A), and from $-0.7\sigma_R$ to $-0.5\sigma_R$ (CEP-1718B).

\section{Characteristics of solutions for Cepheids in DLEBs in final model grid}\label{secapp:tabsfinal}

In this Appendix we present parameters of the best-matching solutions for CEP-0227 (Tab.~\ref{tab:sol-0227}, displayed in Fig.~\ref{fig:227_2}),  CEP-4506 (Tab.~\ref{tab:sol-4506}, displayed in Fig.~\ref{fig:4506_2}),  CEP-2532 (Tab.~\ref{tab:sol-2532}, displayed in Fig.~\ref{fig:2532_2}), and  CEP-1718 (Tab.~\ref{tab:sol-1718}, displayed in Fig.~\ref{fig:1718_2}).

\begin{table}
\caption{Parameters of the best-matching models for CEP-0227, displayed in Fig.~\ref{fig:227_2}.}
\label{tab:sol-0227}
\centering
\begin{tabular}{lrrrr}
\hline\hline
parameter     & primary & note            & secondary & note \\ 
\hline
$M/\MS$       & 4.1732  & $+0.8 \sigma_M$  & 4.0875    & $+0.9\sigma_M$ \\
$q$           & 0.979   & $+0.2\sigma_q$  &           & \\
\Teff{} (K)   & 6098    &                 & 5139      & \\
$\log L/\LS$  & 3.152   &                 & 3.061     & \\
$R/\RS$       & 33.77   & $-9.1 \sigma_R$ & 42.81     & $-14.1 \sigma_R$ \\
$P$ (d)       & 3.79991 & $+0.1 \sigma_P$ &           & \\
Age (Myr)     &  151.2  &                 & 157.3     & \\
$Y_{\rm C}$   & 0.296   &                 & 0.413     & \\
\hline
\end{tabular}
\end{table}

\begin{table}
\caption{Parameters of the best-matching models for CEP-4506, displayed in Fig.~\ref{fig:4506_2}.}
\label{tab:sol-4506}
\centering
\begin{tabular}{lrrrr}
\hline\hline
parameter     & primary & note              & secondary & note \\ 
\hline
$M/\MS$       & 3.6164  & $+0.2 \sigma_M$   & 3.5335    & $+0.5\sigma_M$ \\
$q$           & 0.977   & $+0.7\sigma_q$    &           & \\
\Teff{} (K)   & 6175    &                   & 6047      & \\
$\log L/\LS$  & 3.01    &                   & 2.92      & \\
$R/\RS$       & 27.95   & $-2.8 \sigma_R$   & 26.39     & $-14.1 \sigma_R$ \\
$P$ (d)       & 2.98888 & $+0.0 \sigma_P$   &           & \\
Age (Myr)     & 201.2   &                   & 207.3     & \\
$Y_{\rm C}$   & 0.062   &                   & 0.335     & \\
\hline
\end{tabular}
\end{table}

\begin{table}
\caption{Parameters of the best-matching models for CEP-2532, displayed in Fig.~\ref{fig:2532_2}.}
\label{tab:sol-2532}
\centering
\begin{tabular}{lrrrr}
\hline\hline
parameter     & primary & note              & secondary & note \\ 
\hline
$M/\MS$       & 3.9752  & $-0.0\sigma_M$    & 3.9210    & $-0.2\sigma_M$ \\
$q$           & 0.986   & $-0.5\sigma_q$    &           & \\
\Teff{} (K)   & 6397    &                   & 4805      & \\
$\log L/\LS$  & 3.072   &                   & 2.846     & \\
$R/\RS$       & 27.98   & $-0.9\sigma_R$    & 38.20     & $+0.1\sigma_R$ \\
$P$ (d)       & 2.03603 & $+0.0\sigma_P$    &           & \\
Age (Myr)     & 175.3   &                   & 174.9     & \\
$Y_{\rm C}$   & 0.194   &                   & 0.589     & \\
\hline
\end{tabular}
\end{table}

\begin{table}
\caption{Parameters of the best-matching models for CEP-1718, displayed in Fig.~\ref{fig:1718_2}.}
\label{tab:sol-1718}
\centering
\begin{tabular}{lrrrr}
\hline\hline
parameter     & primary & note              & secondary & note \\ 
\hline
$M/\MS$       & 4.2296  & $-1.0\sigma_M$    & 4.2547    & $+0.9\sigma_M$ \\
$q$           & 1.006   &   $+3.6\sigma_q$  &           & \\
\Teff{} (K)   & 6518    &                   & 6198      & \\
$\log L/\LS$  & 3.109   &                   & 3.140     & \\
$R/\RS$       & 28.10   & $+0.3\sigma_R$    & 32.24     & $-0.7\sigma_R$ \\
$P$ (d)       & 1.96829 & $+0.2\sigma_P$    & 2.47962 & $-0.1\sigma_P$\\
Age (Myr)     & 152.7   &                   & 153.3     & \\
$Y_{\rm C}$   & 0.394   &                   & 0.181     & \\
\hline
\end{tabular}
\end{table}

\FloatBarrier

\section{Additional solutions for Cepheids in DLEBs}\label{secapp:addsol}

\begin{figure}
\centering
\includegraphics[width=\linewidth]{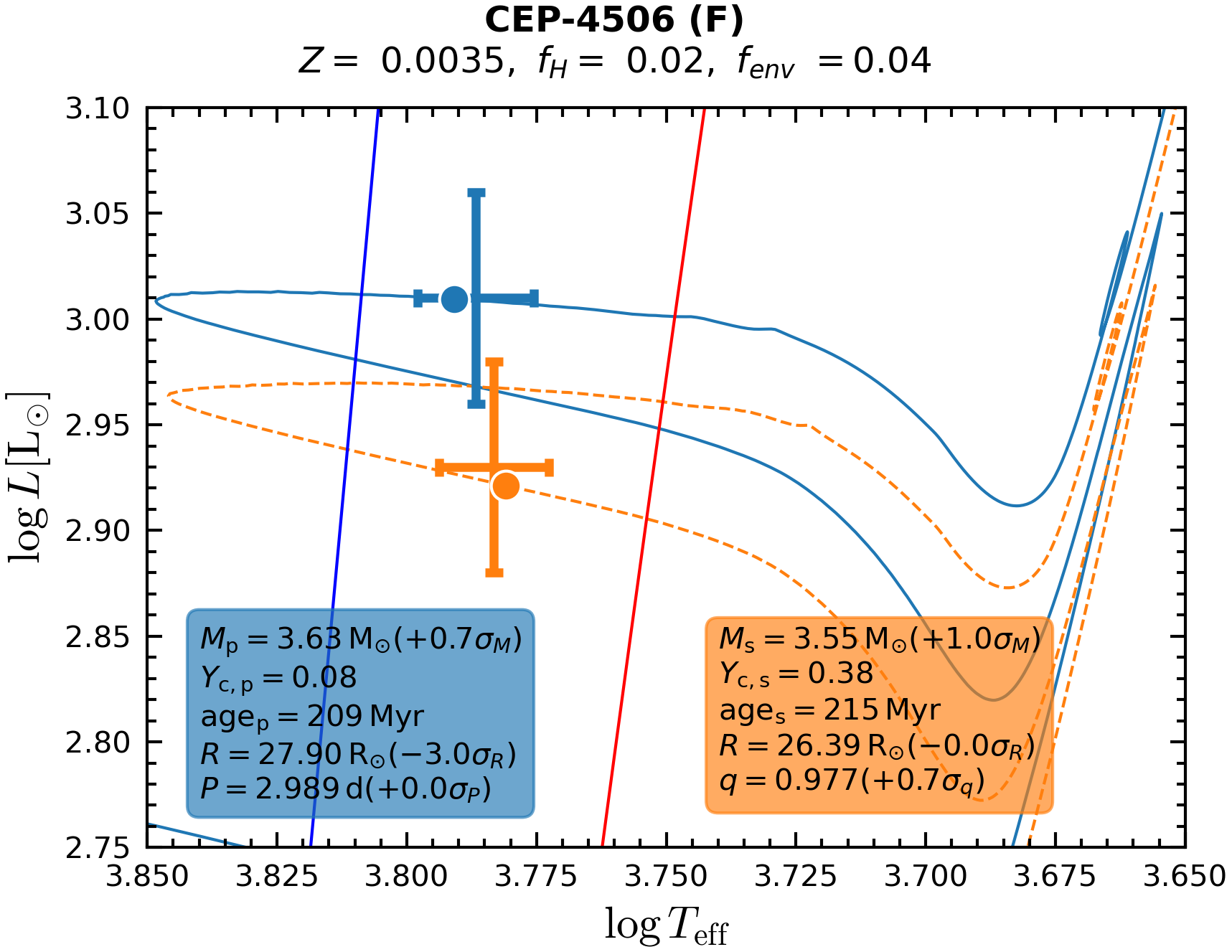}
\caption{Alternative solution in final grid for CEP-4506.}
\label{fig:4506_2b}
\end{figure}

\begin{figure}
\centering
\includegraphics[width=\linewidth]{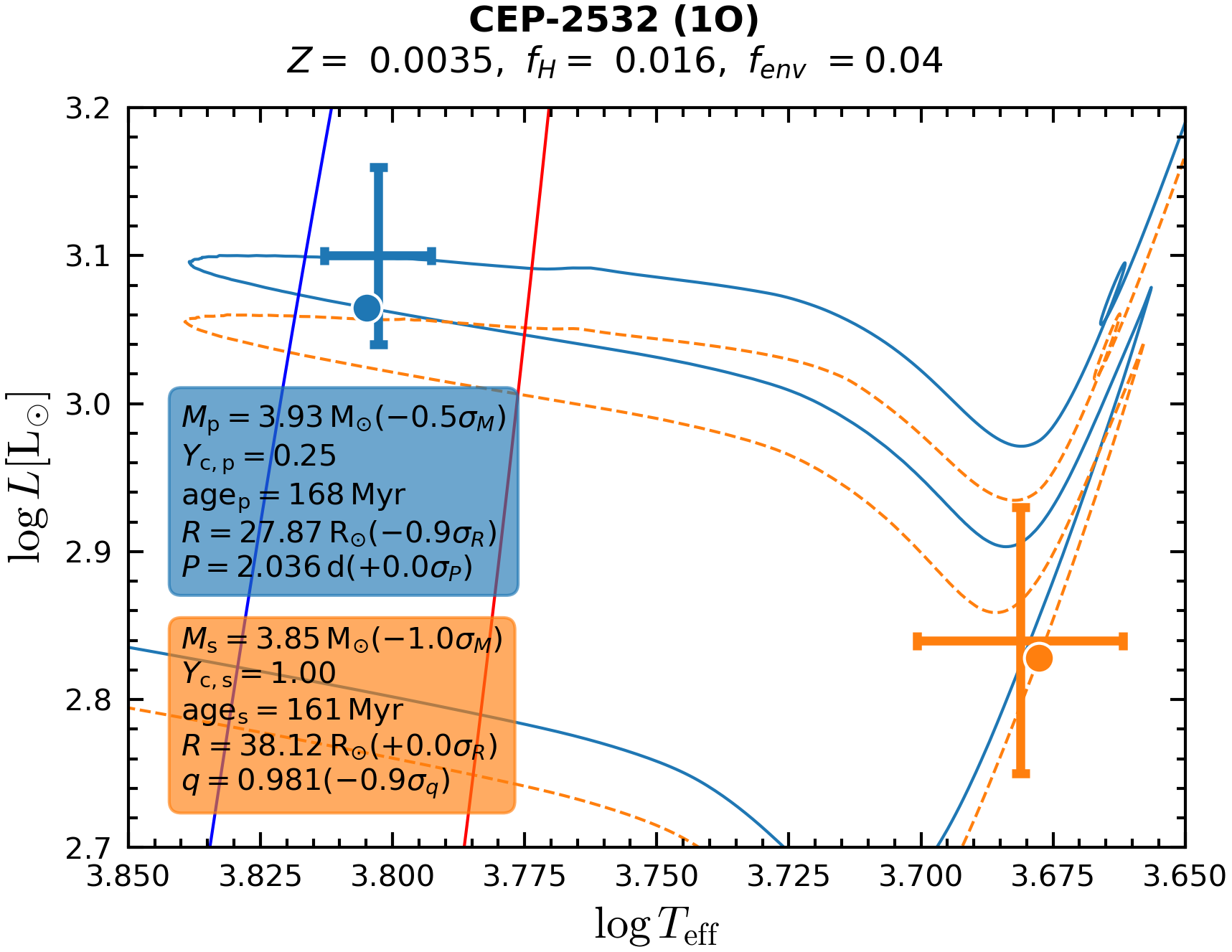}
\caption{Alternative solution in final grid for CEP-2532.}
\label{fig:2532_2b}
\end{figure}

\begin{figure}
\centering
\includegraphics[width=\linewidth]{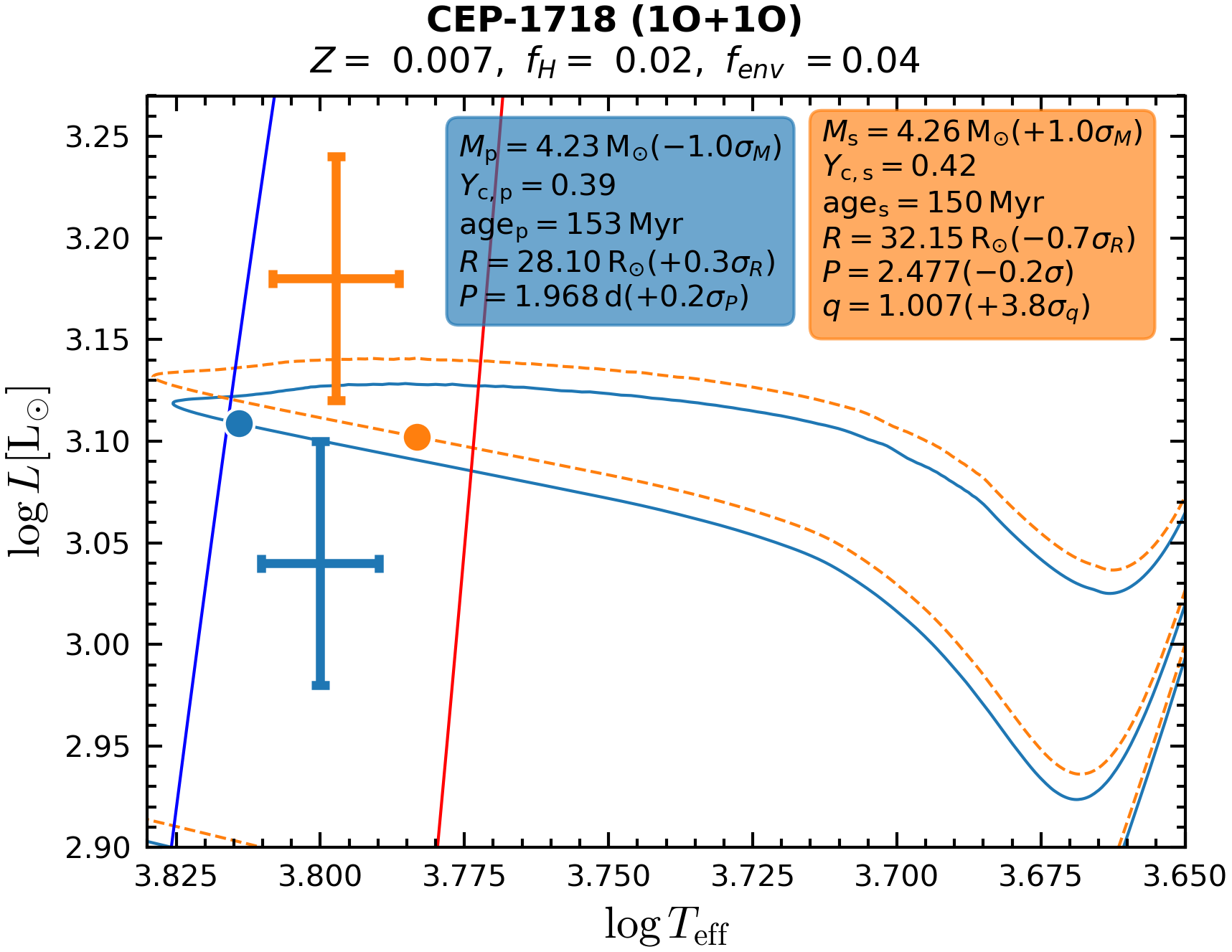}
\caption{Alternative solution in final grid for CEP-1718, with secondary forced on the second crossing in accordance with PCR measured by \citetalias{Pilecki-2018}.}
\label{fig:1718_2b}
\end{figure}

In this Appendix we present additional solutions for some of the modeled DLEBs, CEP-4506 (Fig.~\ref{fig:4506_2b}), CEP-2532 (Fig.~\ref{fig:2532_2b}), and CEP-1718 (Fig.~\ref{fig:1718_2b}), along with their numerical characteristics in Tabs ~\ref{tab:sol-4506_alt}, \ref{tab:sol-2532_alt} and \ref{tab:sol-1718_alt}, respectively.

\begin{table}
\caption{Parameters of an alternative solution for CEP-4506, displayed in Fig.~\ref{fig:4506_2b}.}
\label{tab:sol-4506_alt}
\centering
\begin{tabular}{lrrrr}
\hline\hline
parameter     & primary & note              & secondary & note \\ 
\hline
$M/\MS$       & 3.6318  & $+0.7\sigma_M$   & 3.5488   & $+1.0\sigma_M$ \\
$q$           & 0.977   & $+0.7\sigma_q$    &           & \\
\Teff{} (K)   & 6179    &                   & 6038      & \\
$\log L/\LS$  & 3.01    &                   & 2.92      & \\
$R/\RS$       & 27.90   & $-3\sigma_R$      & 26.39     & $-0.0\sigma_R$ \\
$P$ (d)       & 2.98918 & $+0.0\sigma_P$    &           & \\
Age (Myr)     & 208.7   &                   & 215.3     & \\
$Y_{\rm C}$   & 0.080   &                   & 0.377     & \\
\hline
\end{tabular}
\end{table}

\begin{table}
\caption{Parameters of an alternative solution for CEP-2532, displayed in Fig.~\ref{fig:2532_2b}.}
\label{tab:sol-2532_alt}
\centering
\begin{tabular}{lrrrr}
\hline\hline
parameter     & primary & note              & secondary & note \\ 
\hline
$M/\MS$       & 3.9252  & $-0.5\sigma_M$    & 3.8498   & $-1.0\sigma_M$ \\
$q$           & 0.981   & $-0.9\sigma_q$    &           & \\
\Teff{} (K)   & 6382    &                   & 4761     & \\
$\log L/\LS$  & 3.065   &                   & 2.828      & \\
$R/\RS$       & 27.87   & $-0.9\sigma_R$    & 38.12     & $+0.0\sigma_R$ \\
$P$ (d)       & 2.03585 & $+0.0\sigma_P$    &           & \\
Age (Myr)     & 167.9   &                   & 160.5     & \\
$Y_{\rm C}$   & 0.248   &                   & 0.997     & \\
\hline
\end{tabular}
\end{table}

\begin{table}
\caption{Parameters of an alternative solution for CEP-1718, displayed in Fig.~\ref{fig:1718_2b}.}
\label{tab:sol-1718_alt}
\centering
\begin{tabular}{lrrrr}
\hline\hline
parameter     & primary & note              & secondary & note \\ 
\hline
$M/\MS$       & 4.2296  & $-1.0\sigma_M$    & 4.2601    & $+1.0\sigma_M$ \\
$q$           & 1.007   & $+3.8\sigma_q$    &           & \\
\Teff{} (K)   & 6518    &                   & 6071      & \\
$\log L/\LS$  & 3.109   &                   & 3.102     & \\
$R/\RS$       & 28.10   & $+0.3\sigma_R$    & 32.15     & $-0.7\sigma_R$ \\
$P$ (d)       & 1.96829 & $+0.2\sigma_P$    & 2.47700   & $-0.2\sigma_P$ \\
Age (Myr)     & 152.7   &                   & 149.7     & \\
$Y_{\rm C}$   & 0.395   &                   & 0.420     & \\
\hline
\end{tabular}
\end{table}

\section{Period--radius relations and the nonlinear radius shift}\label{secapp:pr}

\begin{figure}
\centering
\includegraphics[width=\linewidth]{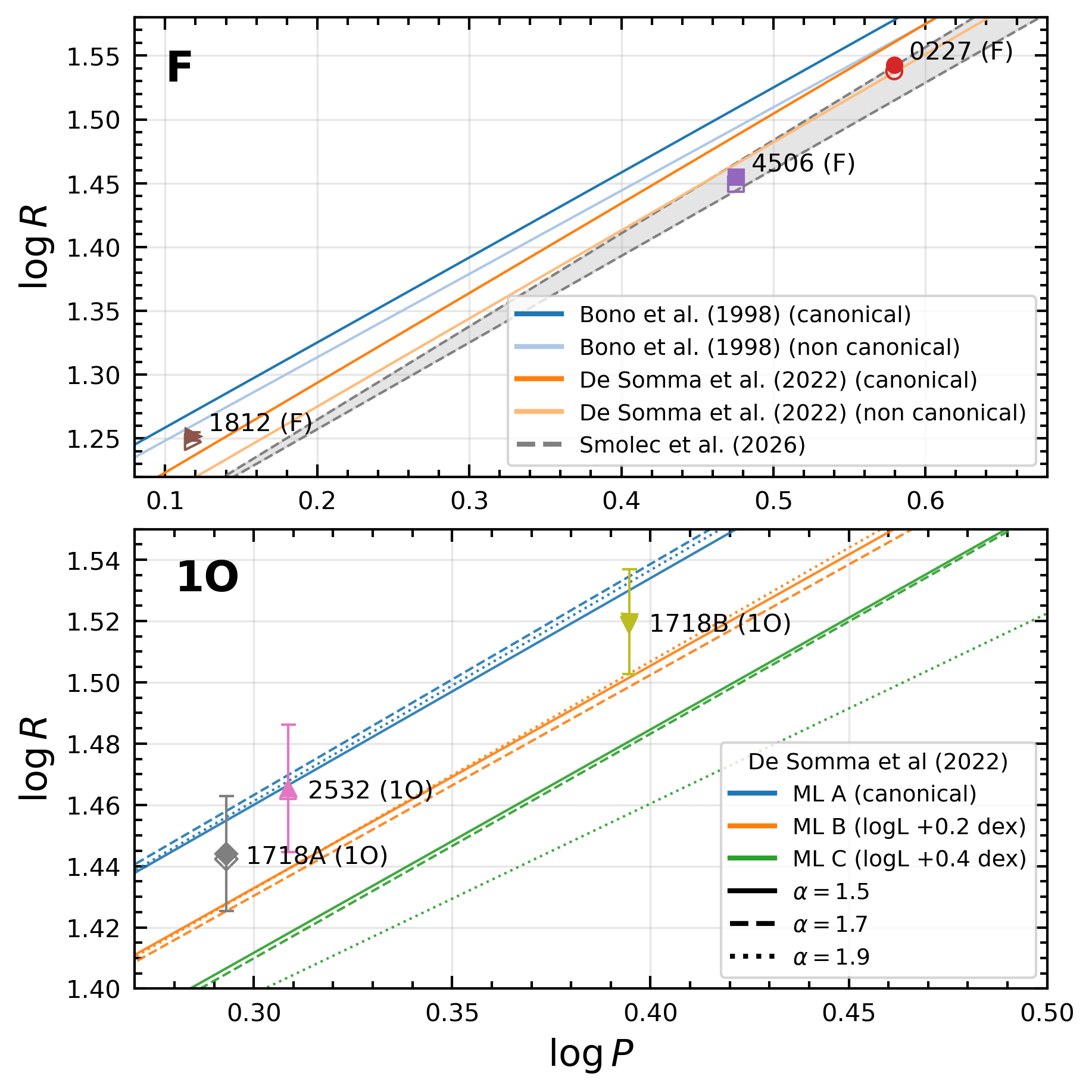}
\caption{Period-radius relations from various sources \citep{Bono-1998, DeSomma-2022, Smolec-2026}, as indicated in the legends, confronted with radius determinations of considered Cepheids in eclipsing binary systems \citep{Pilecki-2018}. Open symbols illustrate the expected effect of nonlinear radius shift.}
\label{figapp:pr}
\end{figure}

In Fig.~\ref{figapp:pr} we confront the radius determinations for the considered systems with predictions of several PR relations based on both purely static evolutionary models and nonlinear pulsation models. The latter, by construction, include the nonlinear radius correction, while the former do not. Fundamental-mode (F) and first-overtone (1O) Cepheids, together with the corresponding relations, are plotted separately in the top and bottom panels of the figure. The order-of-magnitude higher precision of the radius determinations for F-mode Cepheids is immediately visible, as the error bars are essentially smaller than the point size. Open symbols indicate the expected size of the nonlinear radius shift based on the data in Tab.~\ref{tab:P}, which is also very small.

Considering the PR relations shown in the top panel, the gray dashed lines correspond to the relation based on evolutionary models including MS core overshooting that we computed in \cite{Smolec-2026} (based on the O24 model set for the second crossing, assuming LMC metallicity). For all relations involving pulsation period, a significant color term is expected. The two dashed lines illustrate this point: the upper relation is computed along the blue edge of the instability strip, while the lower one is computed along the red edge. At a given period, the color effect is significantly larger than both the precision of the radius determinations for the considered stars and the expected nonlinear radius shift. In the literature, however, PR relations are most often given without an explicit color term. The two other families of PR relations plotted in Fig.~\ref{figapp:pr} are taken from \cite{Bono-1998} (from their Tab.~1) and \cite{DeSomma-2022} (from their Tab.~3). Both are based on nonlinear pulsation models, either canonical (without overshooting, mass loss, or rotation) or non-canonical (including the effects of MS core overshooting). While these relations predict larger radii at a given pulsation period than the relation based on evolutionary models, this difference is largely due to the different assumptions underlying the calculations and not to the nonlinear radius shift, which is a very small effect. The relations themselves differ considerably depending on the adopted mass--luminosity relation (i.e., whether MS core overshooting is included) and on whether they are based on the older models of \cite{Bono-1998} or the updated calculations of \cite{DeSomma-2022}. Comparing the relations with the locations of the sample Cepheids, it is difficult to draw any firm conclusions, as the relations lack a color term and assume specific metallicities and overshooting efficiencies (or, equivalently, specific underlying mass--luminosity relations). Clearly, we cannot conclude that the observations favor PR relations based on nonlinear pulsation models, simply because the nonlinear radius shift is a much smaller effect than the shifts in the PR relation caused by color, metallicity, or the adopted mass--luminosity relation.

In the bottom panel of Fig.~\ref{figapp:pr} we show the 1O Cepheids together with the corresponding PR relations from \cite{DeSomma-2022}, computed assuming three different mass--luminosity relations as well as different mixing-length parameters, $\alpha$ (from their Tab.~4). Again, it is clear that the shifts caused by the adopted mass--luminosity relation are significantly larger than both the expected nonlinear radius shifts and the shifts in the PR relations arising from different choices of the mixing-length parameter. The latter two effects are of comparable magnitude.

This illustrates that while PR relations are very useful for general studies and comparisons involving large samples of Cepheids, and may indeed serve as valuable probes of stellar evolution and pulsation theory, including the underlying mass--luminosity relation and related assumptions \citep[see, e.g.,][]{Anderson-2016, DeSomma-2022, Smolec-2026}, they are not well suited to probe the effect discussed in this paper, namely the nonlinear radius shift. To do so, one would need a self-consistent set of PR relations based on evolutionary and pulsation models, ideally computed with the same evolutionary and pulsation code. In this context, we note that \MESA{} now allows fully consistent pulsation calculations on top of evolutionary models \citep[see][]{Farag-2026}. Even then, however, the effects of color and metallicity dependence as well as the effects of overshooting, the adopted convective model parameters \citep[see][]{DeSomma-2022}, and other mixing processes would remain much larger than the nonlinear radius shift itself. To study this effect, detailed modeling of individual stars is required.

\clearpage

\end{appendix}
\end{document}